\documentclass[reprint,amsmath,amssymb,aps]{revtex4-1}

\usepackage{graphicx}
\usepackage{dcolumn}
\usepackage{bm}
\usepackage{color}
\usepackage{amsmath}
\usepackage{colortbl}
\usepackage[normalem]{ulem} 
\newcolumntype{M}[1]{>{\centering\arraybackslash}m{#1}}
\newcolumntype{N}{@{}m{0pt}@{}}
\usepackage[justification=centering]{caption}
\begin{document}
\bibliographystyle{elsarticle-num}

\title{Mean-field theory of superradiant phase transition in complex networks}
\author{Andrei Yu. Bazhenov}
\author{Dmitriy V. Tsarev}
\author{Alexander P. Alodjants}%
\affiliation{%
National Research University for Information Technology,
Mechanics and Optics (ITMO), St. Petersburg, 197101, Russia
}%

\date{\today}

\begin{abstract}
In this work we consider a superradiant phase transition problem for the Dicke-Ising model, which generalizes the Dicke and Ising models for annealed complex networks presuming spin-spin interaction. The model accounts for the interaction between a spin-1/2  (two-level) system and external classical (magnetic) and quantized (transverse) fields. We examine regular, random, and scale-free network structures characterized by the $\delta$ function, random (Poisson), and power-law exponent [$p(k) \propto k^{-\gamma}$] degree distributions, respectively. To describe paramagnetic (PM) - ferromagrenic (FM) and superradiant (SR) phase transitions we introduce two order parameters: the total weighted spin $z$-component and the normalized transverse field amplitude, which correspond to the spontaneous magnetization in $z$- and $x$-directions, respectively. 
For the regular networks and vanishing external field we demonstrate that these phase transitions generally represent prerequisites for the crossover from a disordered  spin state to the ordered one inherent to the FM and/or SR phase. 
Due to the interplay between the spin interaction and the finite size effects in  networks we  elucidate novel features of the SR state in the presence of the PM-FM phase transition. In particular, we show that the critical temperature may by high enough and essentially depends on  parameters  which characterize statistical properties of the network structure. 
For the scale-free networks we demonstrate that the network architecture, characterized by the particular value of $\gamma$, plays a key role in the SR phase transition problem. Within the anomalous regime scale-free networks possess a strong effective spin-spin interaction supporting  fully ordered FM state, which is practically non-sensitive to variations of the quantum transverse field or moderate classical magnetic field. In a scale-free regime the networks exhibit vanishing of the collective spin component in $z$-direction with increasing $\gamma$ accompanied by establishing spontaneous magnetization in the transverse field. The SR phase transition occurs in the presence of some FM state.
We establish the conditions for the network parameters,  classical and quantum field features   to obtain a quantum phase transition in the spin system when the critical temperature approaches zero.

\end{abstract}

\maketitle

\section{Introduction}

Currently, complex networks evoke an enormously increasing interest among the scientific community performing advanced studies at the boundary of physics, social and cognitive sciences, and applied mathematics~\cite{Dorogov1,New1,Barabook}. In particular, these studies aim to investigate complex processes in social networks and on the Internet~\cite{Fortunato,Easley}, in living systems integrated within  biological networks ~\cite{Napoli1}, designing new (quantum) materials with complex topology and structure~\cite{Ahnert,Bianconi, Nori}, developing communications and information network technologies including quantum networks~\cite{Barrat,Brito}. Complex networks and, especially, (hyper)network structures are inherent to modern cognitive science and current brain activity research~\cite{Bullmore}. 

Despite the fact complex network studies are largely interdisciplinary, in many cases they are based on the models and approaches of statistical physics, which allow obtaining sufficiently clear dependencies for non-trivial processes in various network structures~\cite{Albert}. The Ising model is one that has been well-established in such works~\cite{Dorogovtsev}. In particular, an Ising-type model was proposed to explain the opinion formation and social impact~\cite{Holyst}. This model was recently revised in the framework of collective emotions~\cite{Tsarev}, social cohesion, and structural balance~\cite{Thurner}.

The Ising model for scale-free networks possessing arbitrary degree distribution was comprehensively studied by Leone et al. in~\cite{Leone} by the replica method. The random transverse Ising model on the complex networks with a scale-free degree distribution was examined in the framework of superconductor-insulator phase transitions~\cite{Bianconi2}. Notably, the annealed network approximation was explored in~\cite{Krasnytska} to characterize scale-free networks.

It seems important to stress that the Ising model allows depicting complex (scale-free) network models exhibiting a second order phase transition and Bose-Einstein condensation (BEC) phenomenon~\cite{Barabasi}. In contrast to random networks, the scale-free, so-called Barab\'asi-Albert (BA), model considers a preferential link connection during network growing. This situation can be established in the framework of the Ising annealed network approach that possesses a power law distribution of degrees~\cite{Bianconi3,Suchecki}. Some peculiarities of a ferromagnetic (FM) phase transition and criticality in such a model, which occurs for large but finite size network systems, represent a primary interest in~\cite{Dorgov2,Aleksiejuk}. The mean-field  approach to 
the Ising model on networks with a degree distribution is discussed in ~\cite{Dorogovtsev}. In recent research ~\cite{Napoli2} Krishnan et al. examined the mean-field approach based on the effective long-range interacting homogeneous Ising model to describe the Ising model on a BA network. By means of Monte-Carlo simulations and analytical approach to BA network it was shown  that  such  a  model is reasonable in a low temperature domain and above the critical temperature.   However, the role of  degree exponent $\gamma$ for scale-free networks was not discussed in  ~\cite{Napoli2}.

Noteworthy, the critical behaviour of the Ising model manifests due to the spin-spin interaction. The interaction of spins with a classical (constant transverse) external field is typically studied in the framework of the so-called transverse Ising model, cf.~\cite{Suzuki}. 

In this work we focus on the problem when the transverse field represents some variable which allows for a second quantization procedure. In this case, spin systems may be represented as two-level oscillators like natural or artificial two-level atoms~\cite{Nori2}, quantum dots, etc, which interact with the quantized field in the framework of the Dicke model. This model presumes a so-called superradiant (SR) second order phase transition, and has been long known in quantum optics domain~\cite{Hepp,Wang,Emary,Larson}. In particular, the SR phase transition evokes the establishment of some certain (non-zero) spontaneous polarization, that occurs in a thermodynamically equilibrium ensemble of two-level oscillators interacting with the quantized field. 

Up to present, the SR phase transition has been predicted and observed with atomic ensembles~\cite{Bohnet, Chestnov, Akkermans}, exciton polaritons in semiconductor microstructures~\cite{Eastham}, superconductor circuits~\cite{Wang2,Bamba}, solids~\cite{Cong}, and extended star graphs~\cite{Yalouz}. The evidence of the SR state in such experiments is usually achieved by means of cavity exploiting that enables to enhance a photon lifetime~\cite{Nori,Vahala,Al1}. Moreover, the quantized light interaction with various two-level system ensembles  enables  high (up to room) temperature phase transitions, cf.~\cite{Eastham, Chestnov}.  Such a feature of the Dicke-Ising model that we offer in this work,  may be useful for the problem of obtaining high-temperature  BEC, which is now a hot topic of research in statistical physics and material science, cf. ~\cite{Yamamoto, Chest1}.  

In particular, two different regimes are relevant to be distinguished when the SR phase transition occurs in the system. First, we speak about the zero chemical potential limit when we can describe the total system in the framework of the canonical ensemble approach. In this case, the ensemble of two-level oscillators and photonic system are thermodynamically equilibrium closed sub-systems~\cite{Hepp,Wang,Emary,Larson}. Second, the grand canonical approach presumes a non-zero chemical potential when photons and spins (or two-level systems) can form coupled states (dressed states, polaritons), which possess the SR phase transition~\cite{Eastham, Chestnov}. In other words, at non-zero chemical potential we can speak about BEC of low branch polaritons that occurs in the system in the presence of appropriate trapping potential ~\cite{Yamamoto, Ber, Chest1}. In this work we restrict ourselves by  canonical ensemble approach to the network  spin-1/2 system, which interacts with classical and quantized fields.

Recently, the SR phase transition has been discussed in~\cite{Gammelmark,Lee} in the framework of the so-called Dicke–Ising model for the material systems exhibiting spin-spin interaction. In particular, in~\cite{Gammelmark,Lee} it is demonstrated that such systems permit a first order phase transition depending on the character of spin-spin interaction. Some important applications in quantum metrology are also found~\cite{Bazhenov}.
 
In~\cite{Tsarev} we established the Dicke-like model for the collective decision-making problem that exhibits the second order phase-transition that occurs in heterogeneous information-oriented communities interacting with information field. In particular, we showed that the system demonstrates social polarization and lasing phenomena for certain parameters (density of excitations, temperature). In this sense our model bridges the gap between the laser-like models described in~\cite{Khrennik1, Khrennik2} and current studies on opinion formation which involve echo-chamber effects in social network communities~\cite{Cota2, Baumann1}. However, the  network topology and (Ising-like) coupling between agents  eventually play a vital role for various socially oriented statistical models, see \cite{Fortunato,Easley, Bianconi3,Suchecki, Cota2, Baumann1, Bukh} and cf.~\cite{Tsarev}.

Surprisingly, the phase transitions problem for the Dicke–Ising model specified for complex networks has not been studied yet at all; this work aims to investigate it. We are going to elucidate the role of particular network characteristics, such as node degree, degree exponent (for scale-free networks) in the SR phase transition that activates the network structure. In particular, an important task of this work is to show how the establishment of the superradiant transverse quantum field affects the spin ordering in the orthogonal $z$-direction.

The paper is arranged as follows. In Sec.~II we offer the Dicke-Ising model that exploits the annealed complex networks approach. The equations for the order parameters are derived in the framework of variational (thermodynamic) approach. Specifically, we explore the mean-field approach, which is familiar in quantum optics, cf.~\cite{Wang,Eastham,Al1}. This approach deals with coherent state anzatz for quantized field, which presumes neglect of spin-spin and spin-quantized field  correlations. However, as we show, it allows to account degree correlations in the network structure.  In Sec.~III we discuss the complex network parameters that play an essential role in the phase transition problem. Moreover, we examine regular, random, and scale-free networks. The scale-free networks are examined in the anomalous, scale-free, and random regimes, which are characterized by different values of degree exponent, cf.~\cite{Fortunato,Easley}. The phase transition problem for the complex networks in the presence of the quantized transverse field is comprehensively studied in Sections ~IV, V. 
We investigate various limits in respect of the classical magnetic field, degree exponent, statistical properties of network degree. 
In Sec.~IV we examine regular networks characterized by a constant spin-spin interaction strength. In this case, we obtain simple analytical treatments describing the SR phase transition in the presence of FM and/or paramagnetic (PM) states for both low and high temperature limits. Sec.~V presents the solution of the phase transition problem in the scale-free and random networks, which possess strong spin-spin interaction depending on the network architecture. In addition, we study the quantum phase transition problem for the complex networks attained in the zero temperature limit. Some specific problems connected with the SR phase transition in the low temperature limit with regular networks are given in Appendix A.
In Sec.~VI the results obtained are summarized. 

\section{The Dicke-Ising model for complex networks}

Let us consider the ensemble of $N$ spin-1/2
(or two-level) systems (particles), which randomly occupy $N$ nodes of a complex network. We represent the complex network as a graph with non-trivial (specific) properties, resulting from its topology, degree distribution, and other characteristics, see~\cite{Dorogov1,New1,Easley}. There are spins, which
are placed in the nodes of the graph, are supposed  to interact with classical (local) magnetic field $h_i$ and the quantized (transverse) field. We describe the transverse field by means of annihilation ($a$) and creation (${a}^\dag$) operators. The total Hamiltonian of the model reads 

\begin{widetext}
\begin{eqnarray}\label{Hamiltonian} 
\textrm{H}=-\sum\limits_{ij} {{J_{ij}}\sigma^{z}_i\sigma^{z}_j}-\frac{1}{2}\sum\limits_{i} {h_i\sigma^{z}_i}+\omega_aa^{\dag}a-\frac{1}{2\sqrt{N}}\sum\limits_{i} {\chi_i\sigma^{x}_i(a+a^{\dag})},
\end{eqnarray}
\end{widetext}
where $\sigma^{z}_i$ and $\sigma^{x}_i$, $i=1,...,N$, characterize the $i$-th particle spin components in $z$- and $x$-directions, respectively. The sum is performed over the graph vertices with certain adjacency matrix $A_{ij}$ proportional to $J_{ij}$, it stores the information about the graph structure: matrix element $A_{ij}=1$ if two vertices are linked and $A_{ij}=0$ otherwise. 

First two terms in Eq.~\eqref{Hamiltonian} describe the Ising model, while the last terms are inherent to the Dicke model. The Dicke part of \textrm{H} in Eq.~\eqref{Hamiltonian} is responsible for the spin interaction with the quantized transverse ("photonic") field possessing energy $\hbar\omega_a$ (in this work for simplicity we put the Planck and Boltzmann constants $\hbar=1$, $k_B=1$). Parameter $\chi_i$ in~\eqref{Hamiltonian} characterizes the coupling of spin $x$-component, $\sigma^{x}_i$, with the quantum transverse field in the so-called dipole approximation. Below we restrict ourselves to a homogeneous problem when $h_i=h$ and $\chi_i=\chi$ for any $i$. 
 
Notably,~\eqref{Hamiltonian} describes the Husimi-Temperley-Curie-Weiss model that belongs to the transverse Ising model if in~\eqref{Hamiltonian} we assume $\chi_i(a+{a}^\dag)/\sqrt{N} \to \epsilon_i$, i.e. in the limit of the fixed, constant, classical transverse field, cf.~\cite{Suzuki}. However, in this work average photon number $N_{ph}\equiv{\langle}a^{\dag}a{\rangle}$ accumulated in the transverse field is a variable that relates to the order parameter of the SR phase transition.

We are more interested in the annealed network approach that presumes a weighted, fully connected graph model. This network is dynamically rewired. Two nodes $i$ and $j$ are connected with probability $p_{ij}$ that looks like, cf.~\cite{Bianconi2}
\begin{eqnarray}\label{Prob} 
p_{ij}=P(A_{ij}=1)=k_{i}k_{j}/{N}\langle{k}\rangle,
\end{eqnarray}
where $A_{ij}$ is an element of the adjacency matrix, 
$k_i$ is $i$-th node degree taken from distribution $p(k)$. In~\eqref{Prob} $\langle{k}\rangle={\frac{1}{N}}\sum\limits_{i}{k_i}$ is an average degree.

Noteworthy, the annealed network approach is valid for $p_{ij}\ll1$ and for large enough $N$, cf.~\cite{Lee2}. We recast parameter $J_{ij}$ that indicates the coupling between the nodes in Eq.~\eqref{Hamiltonian} through probability $p_{ij}$ as $J_{ij}=Jp_{ij}$, where $J$ is a constant. 

Thus, the strength of two spins interaction $J_{ij}$ is a variable parameter and depends on particular network characteristics; it is greater for two pairs of nodes with the highest $k$ coefficient.

The properties of the system described by Eq.~\eqref{Hamiltonian} may be determined by means of two order parameters. The first order parameter, $S_z$, is a collective weighted spin component defined as 
\begin{eqnarray}\label{spin} 
S_z=\frac{1}{N{{\langle}k{\rangle}}}\sum\limits_{i}k_i\sigma^{z}_i. 
\end{eqnarray}
The second one, $\lambda$, is a normalized mean transverse field amplitude 
\begin{eqnarray}\label{photon}
\lambda=\frac{|\alpha|}{\sqrt{N}}=\sqrt{\frac{N_{ph}}{N}}, 
\end{eqnarray}
where we use Glauber coherent state basis $|{\alpha}\rangle$ for the quantized transverse field that contains $N_{ph}={\langle}\alpha|a^{\dag}a|{\alpha}{\rangle}=|\alpha|^2$ photons on average.

The physical meaning of Eqs.~\eqref{Prob},~\eqref{spin} becomes more evident if we introduce local  effective magnetic field $H_{eff,i}\equiv 2 \sum\limits_{j} {{J_{ij}}\sigma^{z}_j}+h$ that acts on the $i$-th node spin. Combining the first two terms of~\eqref{Hamiltonian} with~\eqref{Prob} for $H_{eff,i}$ we obtain \begin{eqnarray}\label{eff_mag}
H_{eff,i}=2J k_i S_z + h.
\end{eqnarray}
The first term in $H_{eff,i}$ depends both on spin-spin interaction strength $J$ and collective weighted spin component $S_z$. The network system peculiarities become unimportant in the limit of strong classical field $h\gg J$. On the other hand, for $h=0$ spin ordering completely depends on local properties of the $i$-th node connectivity.

We can take into account the cumulative role of locally acting effective magnetic fields $H_{eff,i}$ by total effective field $H_{eff}=\frac{1}{N}\sum\limits_{i} H_{eff,i}$ that looks like
\begin{eqnarray}\label{eff_mag2}
H_{eff}=2J \langle{k}\rangle S_z +h.
\end{eqnarray}
 From~\eqref{eff_mag2} it is clear that spin-dependent peculiarities, which are determined by $H_{eff}$, depend on the network topology encoded in average degree $\langle{k}\rangle$. Thus, the limit of zero classical field $h=0$ is of primary interest to elucidate these peculiarities.

To determine order parameters~\eqref{spin},~\eqref{photon} for some temperature $T\equiv1/\beta$, we use a variational (thermodynamic) approach, see e.g.~\cite{Hepp}. We exploit partition function $Z=Tr(e^{-{\beta}\textrm{H}})$. The mean-field approach that we use in this work (cf. ~\cite{Wang}),  presumes exploring factorized  total state anzatz $|\Psi\rangle$ and  representing it as $|\Psi\rangle=|{\alpha}\rangle|{s_{1}...s_{N}}\rangle$, where $|{s_{1}...s_{N}}\rangle$ defines an $N$-spin state. Strictly speaking, for the partition function we have:
\begin{eqnarray}\label{Fun2}
Z=\frac{1}{\pi}\sum\limits_{s_i}{\int{d^{2}\alpha}\langle}\Psi|e^{-{\beta}\textrm{H}}|{\Psi}{\rangle}.
\end{eqnarray}
Assuming ${\int{\frac{d^{2}\alpha} {\pi}}}=N \int_{0}^{\infty} d{\lambda^2}$ for the transverse coherent photonic state and performing integration in~\eqref{Fun2}, for partition function $Z$ we obtain 
\begin{eqnarray}\label{Fun3}
\begin{gathered}
Z=N\int{d{\lambda}^2}e^{-\beta{N}\omega_{a}\lambda^2}e^{-{\beta}JN{\langle{k}\rangle}S_z^2}\\
\prod\limits_i{2\textrm{cosh}\left (\frac{\beta}{2}\sqrt{(4J{k_i}{S_z}+h)^2+4\chi^2\lambda^2}\right)}.
\end{gathered}
\end{eqnarray}

Evaluating the integral in Eq.~\eqref{Fun3} by Laplace method after some calculations we obtain the mean-field equations for collective spin $S_z$, Eq.~\eqref{Sadis}, and average transverse field $\lambda$, Eq.~\eqref{lbdis}, respectively: 

\begin{subequations}\label{systdis}
\begin{align}
\begin{split}
S_z=\frac{1}{N{\langle{k}\rangle}}\sum\limits_{i}k_i\frac{{\Theta}S_zk_i+H}{\sqrt{({\Theta}S_zk_i+H)^2+4\lambda^2}}\\
\textrm{tanh}\left(\frac{\beta}{2}\sqrt{({\Theta}S_zk_i+H)^2+4\lambda^2}\right),
\end{split} \label{Sadis}\\
\begin{split}
\lambda {\Omega}_a=\lambda\frac{1}{N}\sum\limits_{i}\frac{\textrm{tanh}(\frac{\beta}{2}\sqrt{({\Theta}S_zk_i+H)^2+4\lambda^2})}{\sqrt{({\Theta}S_zk_i+H)^2+4\lambda^2}},
\end{split} \label{lbdis}
\end{align}
\end{subequations}

In Eqs.~\eqref{systdis} we introduce the normalized dimensionless parameters as 
\begin{eqnarray}\label{normalization}
\Theta=4J/\chi ; \; \; H=h/\chi; \; \; \Omega_a={\omega_a}/\chi; \; \; {\beta}{\chi}\mapsto \beta.
\end{eqnarray}

Noteworthy, the latter expression in~\eqref{normalization} implies dimensionless temperature $T/\chi\mapsto T$ that we use below. 

Since the number of nodes is large enough, $N\gg1$, we are interested in network structures, which admit continuous degree distribution $p(k)$. The transition from the discrete to continuous version of Eq.~\eqref{systdis} may be performed by replacing $\frac{1}{N}\sum\limits_{i}{...} \rightarrow\int\limits^{k_{max}}_{k_{min}}{...p(k)dk}$, where $k_{min}$ and $k_{max}$ are the minimal and maximal values of node degree $k$. In this case, from~\eqref{systdis} we can obtain:
\begin{subequations}\label{S}
\begin{align}
\begin{split}
S_z=\int\limits^{k_{max}}_{k_{min}}{\frac{kp(k)}{\langle{k}\rangle}\frac{{(\Theta}S_zk+H)}\Gamma \textrm{tanh}\left[\frac{{\beta}}{2}\Gamma\right]dk},
\end{split} \label{Sa}\\
\begin{split}
\lambda {\Omega}_a=\lambda \int\limits^{k_{max}}_{k_{min}}{p(k)\frac{\textrm{tanh}[\frac{{\beta}}{2}\Gamma]}{\Gamma}}dk,
\end{split} \label{lb}
\end{align}
\end{subequations}
where we made denotation $\Gamma\equiv\sqrt{({\Theta}S_z k+H)^2+4{\lambda}^2}$.

Let us briefly describe possible phase states, which may appear in the system determined  by Eqs.~\eqref{S}.

Eq.~\eqref{Sa} is inherent to the FM-PM phase transition. For non-vanishing (constant) $\lambda$ this equation may be connected with the superconductor-insulator phase transition problem, cf.~\cite{Bianconi2}.

Eq.~\eqref{lb} looks as a gap equation in the BCS (Bardeen–Cooper–Schrieffer) theory of superconductivity, cf.~\cite{Landau}. In our case, Eq.~\eqref{lb} governs superradiant properties determined by another order parameter $\lambda$. Physically, $\lambda$ is responsible for the temperature-dependent energy  gap (Rabi splitting) that occurs due to the interaction of the two-level system with the quantized field in the framework of the Dicke model, cf.~\cite{Hepp,Wang,Emary,Larson, Chestnov}. 
This work aims to find joint solutions of Eqs.~\eqref{S}, which correspond to the states with $S_{z}\neq0$, $\lambda\neq0$. 

The phase transitions that we consider can be established by means of spontaneous magnetizations along $x$ and $z$ axes, which are defined as $m_{x,z}=Tr[{\sigma}_i^{x,z}e^{-{\beta}\textrm{H}}]/{Z}$, 
see e.g.~\cite{Suzuki}.
Magnetization in $z$-direction coincides with the collective weighted $z$-component of the spin, i.e. $m_{z}=S_z$. Another magnetization component $m_{x}$ is proportional to the order parameter $\lambda$. This magnetization is similar to spontaneous polarization of two-level system that occurs in the framework of conventional Dicke model, cf.~\cite{Tsarev}. 

The non-superradiant  (non-SR), normal
phase corresponds to the trivial solution of Eq.~\eqref{lb} with $\lambda=0$ 
and is characterized by the absence of transverse magnetization, $m_{x}=0$. In this limit only Eq.~\eqref{Sa} solution is responsible for the phase transition from the PM state ($S_z=0$) to the FM state ($S_z\neq0$) in the network system. Notably, this limit for the Dicke-Ising model may be recognized completely  in the Ising model framework, where the phase transition appears only due to finite size effects, cf.~\cite{Dorgov2,Aleksiejuk}.

\begin{figure*}[ht]
\includegraphics[width=\linewidth]{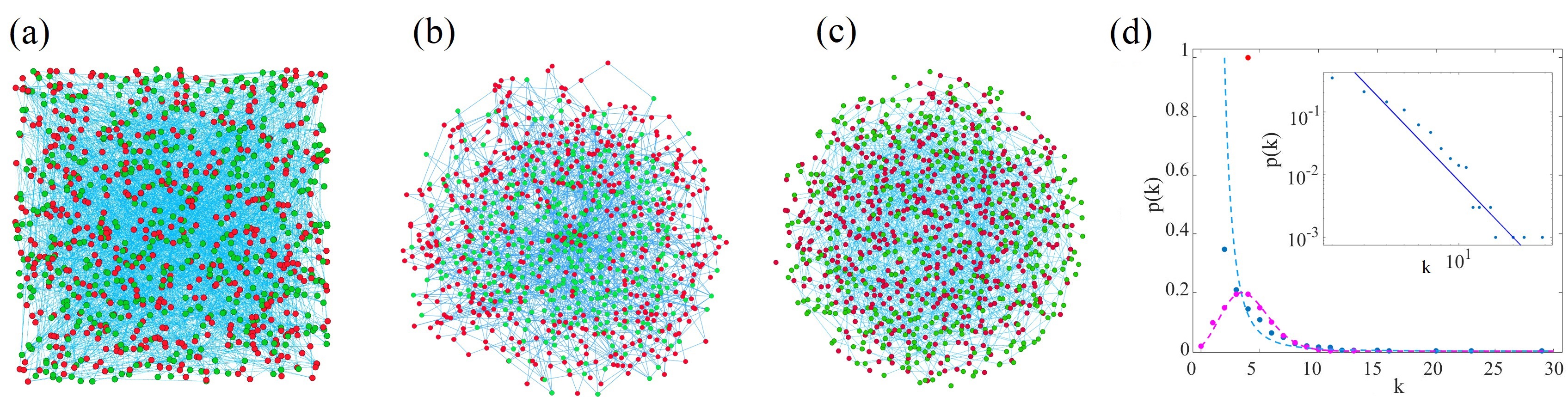}
\caption{(a) Regular,(b) scale-free, (c) random networks  and (d) the relevant degree distributions are plotted for $N=1000$ and ${{\langle}k{\rangle}}=4$. The red (green) nodes in (a)-(c) correspond to the "spin-up" ("spin-down") states. The average magnetization is $S_z=0$. The magenta (blue) curves represent Poisson (power-law) degree distribution in (d). Single (red) point characterizes the Delta-function distribution with $k_0=4$. The insert shows the degree distribution in a logarithmic scale for the scale-free (BA) network.
Points sequences located in right corner in inset, indicate presence of hubs for BA network in (b).}
\label{ris:image}
\end{figure*}

The SR phase corresponds to non-zero (positive) $\lambda$ that characterizes the activation (excitation) of the network nodes possessing spontaneous transverse magnetization $m_{x}\neq0$. The Dicke model is well discussed previously in the framework of SR phase transition~\cite{Hepp,Wang,Emary,Larson}. The model considers non-interacting particles, i.e. we should suppose $J=0$ in \eqref{Hamiltonian}. In this limit Eqs.~\eqref{S} are essentially simplified, and the effects occurring due to network structure properties are completely ignored.

Thus, the transition from some disordered state with $m_{z}=m_{x}=0$ to the ordered one  with $m_{z}\neq0$, and/or $m_{x}\neq0$ appearing for $J\neq0$ represents a primary interest for this work.

\section{Network architectures}

In this work we consider the regular, random, and scale-free networks with the properties determined by various distribution functions $p(k)$.
In particular, we characterize the networks by means of the first (${{\langle}k{\rangle}}$) and second (${{\langle}k^2{\rangle}}$) moments for the degree distribution, which are defined as:
\begin{eqnarray}\label{moments}
{{\langle}k^m{\rangle}}=\int\limits^{k_{max}}_{k_{min}}{{k^m}p(k)dk}, \; \; m=1,2.
\end{eqnarray}

\begin{center}
\begin{table*}[ht]
\caption{Behaviour of average degree ${\langle}k{\rangle}$, $\zeta$-parameter, and critical number $N_c$ for the scale-free network with degree distribution $p(k)\propto k^{-\gamma}$ for various values of degree exponent $\gamma$ in the limit of large $N$. The parameters are specified in the text.}
\label{tabular:Properties}
\begin{tabular}{ | M{4cm} | M{4cm} | M{4cm} | M{4cm} | }
\hline
$\gamma$&${\langle}k{\rangle}$&$\zeta$&${N_c}$\\[6pt] \hline
$\gamma>1$, $\gamma\neq2$,$\gamma\neq3$ &$k_{min}\frac{\gamma-1}{2-\gamma}(N^{\frac{2-\gamma}{\gamma-1}}-1)$&$k_{min}\frac{2-\gamma}{3-\gamma}\frac{N^{\frac{3-\gamma}{\gamma-1}}-1}{N^{\frac{2-\gamma}{\gamma-1}}-1}$&$(\frac{2{\langle}k{\rangle}}{{\Theta}k_{min}^2}\frac{3-\gamma}{\gamma-1}T_c+1)^{\frac{\gamma-1}{3-\gamma}}$\\[6pt] \hline
2&$k_{min}\textrm{ln}(N)$&$\frac{k_{min}}{\textrm{ln}(N)}N$&$\frac{2{\langle}k{\rangle}}{{\Theta}k_{min}^2}T_c$\\[6pt] \hline
3&$2k_{min}$&$\frac{k_{min}}{2}\textrm{ln}(N)$&$e^{\frac{8T_c}{\Theta{\langle}k{\rangle}}}$\\[6pt]
\hline
\end{tabular}
\end{table*}
\end{center}

Below we exploit the parameter
\begin{eqnarray}\label{zeta}
\zeta=\frac{{\langle}k^2{\rangle}}{{\langle}k\rangle} 
\end{eqnarray}
that determines the basic statistical properties of the chosen network.

 The typical networks considered in this work are shown in Fig.~1. Spin-1/2 particles are placed in the network nodes. Fig.1(a)-(c) represents
randomly allocated spin-up and spin-down configuration particles  with the average total spin equal to zero. The configuration  of the network  obeys a certain distribution function, $p(k)$, established in Fig.1(d). In particular, the networks in Fig.1(a)-(c) demonstrate the case where the numbers of spin-up and spin-down particles are equal.

First, we examine a regular network, see Fig.~1(a). It can be established as a network with Delta-function degree distribution $p(k)={\delta}(k-k_0)$ and a certain positive constant degree of nodes $k=k_0$. From definitions~\eqref{moments},~\eqref{zeta} we immediately obtain ${{\langle}k^m{\rangle}}=k_0^m$ ($m=1,2$) and $\zeta=k_0$, respectively. 

Physically, the regular network implies constant interaction strength $\Theta k_{0}$ for an arbitrary pair of spins. It is maximal for the complete graph with $N$ nodes and $k_0=N-1$.

Second, we discuss a scale-free network in the framework of the SR phase transition problem. In particular, we examine the degree distribution obeying the power law 
\begin{eqnarray}\label{powere}
p(k)=\frac{(\gamma-1)k_{min}^{\gamma-1}}{k^{\gamma}},
\end{eqnarray}
where $\gamma$ is a degree exponent, $k_{min}$ is the smallest degree for which Eq.~\eqref{powere} holds. 
In this work we examine the region $1<\gamma\leq4$ which covers the anomalous ($1<\gamma<2$), scale-free ($2<\gamma<3$), and random ($\gamma>3$) regimes, cf.{~\cite{Dorogov1, Barabook}}. The properties of scale-free networks possessing distribution~\eqref{powere} for $\gamma=2$ and $\gamma=3$, should be calculated separately. As an example, in Fig.~1(b) we plot the numerically generated scale-free, Barabási-Albert (BA), network with $\gamma=3$. 

The normalization condition for $p(k)$ is the following: 
\begin{eqnarray}\label{Norma0}
\int\limits^{+\infty}_{k_{min}}{p(k)dk}=1.
\end{eqnarray}

\begin{figure}[h!]
\center{\includegraphics[width=1\linewidth]{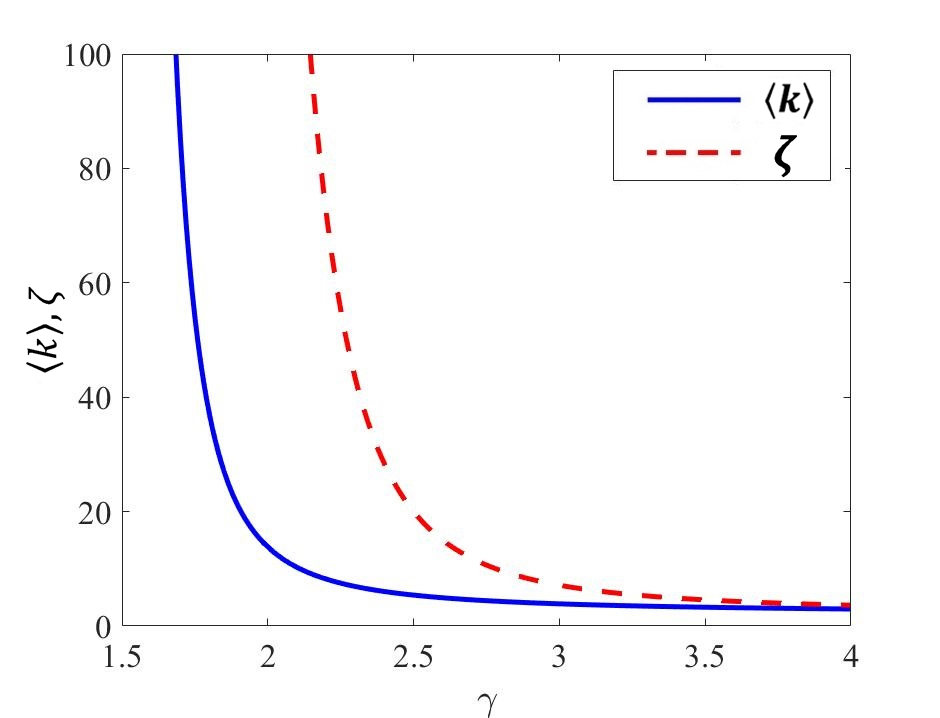}}
\caption{Dependence of mean degree, ${\langle}k{\rangle}$, and $\zeta$-parameter on degree exponent $\gamma$ for the scale-free network at $N=1000$}.
\label{1}
\end{figure}

An important feature of the scale-free network is the existence of hubs, which are clearly recognized by means of three points located in the right corner of the inset in Fig.~1(d). The largest hub is described by degree $k_{max}$ called a natural cutoff. The condition 

\begin{eqnarray}\label{Norma}
\int\limits^{+\infty}_{k_{max}}{p(k)dk}=\frac{1}{N}
\end{eqnarray}
can be used if the network with $N$ nodes possesses more than one node with $k>k_{max}$. From Eqs.~\eqref{powere},~\eqref{Norma} we obtain $k_{max}=k_{min}N^{\frac{1}{\gamma-1}}$, cf. \cite{Barabook}. Notably, in the anomalous regime $k_{max}/k_{min}>N$.

In TABLE I we represent the analytical treatments for the scale-free network characteristics ${\langle}k\rangle$ and $\zeta$ in the limit of large $N$. The relevant dependence of ${\langle}k\rangle$ and $\zeta$ on degree exponent $\gamma$ are presented in Fig.~2. As clearly seen, both characteristics increase in the anomalous region.
It is worth noting that within this region of parameter $\gamma$, the effective magnetic field in Eq. \eqref{eff_mag},~\eqref{eff_mag2} may be enormously large even in the limit of $H=0$. In other words, the networks in anomalous regime support strong spin-spin interaction.

On the other hand, in the scale-free and random regions ${\langle}k\rangle$ and $\zeta$ vanish. As we show below, these features of the scale-free network parameters play a crucial role in the SR phase transition problem.

Third, we consider a random (Poissonian) network model that consists of $N$ nodes and $M$ edges, see Fig.~1(c), cf.~\cite{Dorogov1, Barabook}. Each edge is included in the network  with probability $w$, which is independent from other edges. For a very large $N$ and finite ${{\langle}k\rangle}=(N-1)w\simeq Nw$ it is possible to consider the Poisson degree distribution $p(k)$; it is shown in Fig.~1(d) by the magenda curve: 
\begin{eqnarray}\label{ER_D1}
p(k)=\frac{{{\langle}k\rangle}^k e^{-{\langle}k\rangle}}{k!}.
\end{eqnarray}
From Eqs.~\eqref{moments},~\eqref{zeta}, and~\eqref{ER_D1} we deduce that $\zeta=1+{\langle}k\rangle$ in this case. Noteworthy, the mean-field approach is valid for the Ising model above the critical point ${\langle}k\rangle=1$, when a large cluster with size $N^{2/3}$ is formed~\cite{Barrat}. 

The estimation of upper ($k_{max}$) and lower ($k_{min}$) natural cutoffs for random networks with the Poisson distribution~\eqref{ER_D1} is discussed in~\cite{Barabook}. In particular, we can infer the largest node degree $k_{max}$ of the random network from the numerical solution of $N e^{-{\langle}k\rangle} \frac{{{\langle}k\rangle}^{k_{max}+1}} {(k_{max}+1)!}\approx1$.   This equation may be obtained from the discrete version of \eqref{Norma}. As before we assume that in the random network there exists no more than one node with the degree higher than $k_{max}$. Notably, hubs effectively disappear for the random network since the dependence of $k_{max}$ on $N$ is practically negligible. 

For the random networks, that we examine in this work, average degree $\langle{k}\rangle$ belongs to supercritical regime  domain $1<\langle{k}\rangle<\textrm{ln(N)}$ that is relevant to a moderate number of nodes (we consider networks with  $N=1000$), cf. \cite{Barabook}. Physically, such networks possess isolated nodes, see Fig.1(c). In this regime we can take $k_{min}\geq0$.

\section{Phase transitions in regular networks}

Let us start from the regular network analysis that admits relatively simple analytical solutions of Eqs.~\eqref{S}. Eqs.~\eqref{S} now can be represented as 
\begin{subequations}\label{Reg_syst}
\begin{align}
\begin{split}
S_z=\frac{{\Theta}k_{0}S_z +H}{\Gamma_0}\textrm{tanh}\left[\frac{{\beta}}{2}\Gamma_0\right],\end{split} \label{Reg_Sa}\\
\begin{split}
{\Omega}_a=\frac{\textrm{tanh}[\frac{{\beta}}{2}\Gamma_0]}{\Gamma_0},\end{split}\label{Reg_lb}
\end{align}
\label{system}
\end{subequations}
where $S_z=\frac{1}{N}\sum\limits_{i}{\sigma^{z}_i}$ simply represents the additive (collective) spin $z$-component variable, cf.~\eqref{spin}; $\Gamma_0\equiv\Gamma(k=k_0)=\sqrt{({\Theta}S_z k_0+H)^2+4{\lambda}^2}$. 

Combining together Eqs.~\eqref{Reg_Sa} and~\eqref{Reg_lb} we can establish the necessary condition for the existence of the SR phase transition solution in the system. 

In particular, the solution of~\eqref{Reg_lb} (for FM phase, $S_z\simeq1$) exists if the condition
\begin{eqnarray}\label{cryteria}
({\varepsilon}+\Omega_aH)^2+4\Omega_a^2{\lambda}^2\leq1
\end{eqnarray} 
is fulfilled. In~\eqref{cryteria} we make denotation $\varepsilon={\Theta}k_0{\Omega}_a$. 

When the considered system is very close to the fully ordered FM state, we represent the total spin component as  $S_z\simeq1-\delta$, where $\delta$ is positive small  perturbation ($\delta\ll1$)  to the ordered state. In this case Eqs.\eqref{cryteria} and ~\eqref{Reg_Sa} lead to the condition $2\Omega_a^2{\lambda}^2\leq \delta- 0.5\delta^2$. Notably, the fully ordered FM state  with  $S_z=1$ may be obtained only in the limit of the  absence of any perturbations  ($\delta=0$) that implies  the absence of superradiance,  $\lambda=0$. 

Our approach to the analysis of Eqs.~\eqref{Reg_syst} is as follows.
First, we consider non-zero classical field $H$ that implies some FM state for $S_z$. In this limit it is possible to examine conditions for the occurrence of the SR phase transition.

Second, we examine Eqs.~\eqref{Reg_syst} considering vanishing classical field $H\to 0$, which admits the phase transition from the collective spin disordered state to some ordered one, cf.~\cite{Suzuki}.

\subsection{Ferromagnetic SR phase transition, $H\neq0$}

In the presence of finite magnetic field $H$, some FM phase establishment with non-zero magnetization $S_z$ occurs. 
The SR phase transition boundary can be obtained from Eqs.~\eqref{Reg_syst} by setting $\lambda = 0$. Thus, we obtain

\begin{subequations}\label{Crititcal}
\begin{align}
\begin{split}
S_z=\textrm{tanh}\left[\frac{\beta_c^{(1)}}{2}(\Theta k_{0}S_z+ H)\right],\end{split} \label{Cr_Sa}\\
\begin{split}
{\Omega}_a=\frac{\textrm{tanh}\left[\frac{\beta_c^{(1)}}{2}(\Theta k_{0}S_z +H)\right]}{\Theta k_{0}S_z +H},\end{split}\label{Cr_lb}
\end{align}
\label{system}
\end{subequations}
where 
$\beta_c^{(1)}=1/T_c^{(1)}$ is the reciprocal critical temperature of the SR phase transition in the presence of the FM state.

Further analysis of Eqs.~\eqref{Reg_syst},~\eqref{Crititcal} is easy to perform in two limiting cases for the temperature parameter.

The low temperature limit presumes $\beta\gg1$. In particular, for $\beta\Gamma_0\gg1$, in Eqs.~\eqref{Reg_syst},~\eqref{Crititcal} one can suppose that 
\begin{eqnarray}\label{LowT}
 \textrm{tanh} \left[\frac{{\beta}}{2}\Gamma_0\right]\approx 1- 2e^{-\beta\Gamma_0}. 
\end{eqnarray} 
In this case, we assume that the FM state is fully ordered, i.e. $S_z\simeq1$, and we use Eqs.~\eqref{Reg_lb},~\eqref{Cr_lb} to elucidate the SR phase transition. 

Remarkably, physical temperature $T$ in \eqref{LowT} may by high enough for complete graph possessing large number of nodes $N$ in the low temperature limit implying  $T\ll\Theta N$.   

Critical temperature $T_c^{(1)}$ may be obtained from (20b) by using (21) (see also Appendix A); it is 
\begin{eqnarray}\label{CriticalT}
T_c^{(1)}=\frac{\Gamma_{0,c}}{\textrm{ln}[\frac{2}{1-\Omega_a\Gamma_{0,c}}]}, 
\end{eqnarray} 
where we define $\Gamma_{0,c}\equiv\Gamma_{0}(\lambda=0)=\Theta k_{0} +H$ at the phase transition point $\lambda=0$. 
 
The SR phase transition occurs if $\beta > \beta_c^{(1)}$ or equivalently $T < T_c^{(1)}$.
 
Noteworthy, condition~\eqref{cryteria} at $\lambda = 0$ and Eq.~\eqref{CriticalT} imply the fulfillment of inequality $\Omega_a (\Theta k_{0} +H)\leq1$ that establishes critical maximal degree $k_{c,max}=(\frac{1}{\Omega_a}-H)/\Theta$. The SR phase can exist only for the networks possessing $k_0 \leq k_{c,max}$.
For example, the complete graph inspires critical number of nodes $N_c=\frac{1-H+\Omega_a\Theta}{\Omega_a\Theta}$ obtained from \eqref{cryteria} at $\lambda = 0$. Thereby, the SR state occurs for the networks with $N<N_c$ nodes.

On the other hand, from~\eqref{Crititcal}, and ~\eqref{normalization} at $S_z=1$, $\lambda=0$, we can obtain the critical value $\chi_c$ of coupling strength $\chi$ that looks like 
\begin{eqnarray}\label{coupl}
\chi_c =\sqrt{\omega_a (h+ 4Jk_{0})}. 
\end{eqnarray} 
 
For a given network with some specified value of $k_0$ the SR phase occurs if transverse field coupling parameter $\chi$ obeys condition $\chi <\chi_c$. For vanishing $J$ Eq.~\eqref{coupl} reproduces a well-known result for the Dicke model of superradiance: the critical coupling parameter is $\chi_c=\sqrt{h \omega_a}$, cf.~\cite{Wang,Emary,Larson}. Importantly, in the framework of this model the SR state disappears for zero external field $h=0$ that supports a disordered state with $S_z=0$. 

Let us establish features of the order parameter $\lambda$ in the low temperature domain. From 
Eqs.~\eqref{Reg_lb},~\eqref{Cr_lb} accounting~\eqref{LowT}, we obtain
\begin{eqnarray}\label{orderp}
\lambda \simeq \lambda_0 \sqrt{1-e^{(\beta_c^{(1)}-\beta)\Gamma_{0,c}}},
\end{eqnarray} 
where $\lambda_0$ is the order parameter at temperature $T=0$ ($\beta\to\infty$) - see (A5) in Appendix A. 

In the vicinity of the critical temperature ($\beta\to \beta_c^{(1)}$) 
the behaviour of the order parameter $\lambda$ is reminiscent of familiar temperature dependence $\lambda \propto \sqrt{1-T/T_c^{(1)}}$, which is relevant to SR second order phase transition, cf.~\cite{Chestnov}.

In the high temperature limit, $\beta\ll1$, we can suppose $\textrm{tanh}(\frac{\beta \Gamma_0}{2})\approx \frac{\beta \Gamma_0}{2}$ in Eqs.~\eqref{Reg_syst}. 
Collective spin $S_z$ in this limit approaches
\begin{eqnarray}\label{Tc3}
S_z=\frac{{\beta}H}{2-\Theta k_0 \beta}.
\end{eqnarray}
Eq.~\eqref{Tc3} indicates the reduction of magnetization in $z$-direction with temperature $T$ increasing ($\beta \to 0$) or classical field $H$ suppression. 

\subsection{Phase transitions in the limit of vanishing classical field, $H\to0$}

Here we perform analysis of Eqs.\eqref{Reg_syst} in the PM-FM phase transition domain that accounts for vanishing $S_z$.
To elucidate the phase transition features it is necessary to examine the case of vanishing classical field, $H\to0$, in more detail.
From~\eqref{Reg_syst} we obtain:
\begin{subequations}\label{Heq0}
\begin{align}
\begin{split}
S_z=\frac{{\Theta}k_{0}S_z}{\sqrt{(\Theta k_{0}S_z)^2+4\lambda^2}}\textrm{tanh}\left[\frac{{\beta}}{2}\sqrt{(\Theta k_{0}S_z)^2+4\lambda^2}\right],\end{split} \label{H0_Sa}\\
\begin{split}
 {\Omega}_a=\frac{ \textrm{tanh}[\frac{{\beta}}{2}\sqrt{(\Theta k_{0}S_z)^2+4\lambda^2}]}{\sqrt{(\Theta k_{0}S_z)^2+4\lambda^2}}.\end{split}\label{H0_lb}
\end{align}
\label{system}
\end{subequations}

The Eqs.~\eqref{Heq0} possess the common solution if the condition
\begin{eqnarray}\label{epsi}
\varepsilon\equiv\Theta k_{0}{\Omega}_a=1 
\end{eqnarray} 
is fulfilled, cf.~\eqref{cryteria}. Eqs.~\eqref{H0_Sa} and~\eqref{H0_lb} with condition~\eqref{epsi} are reduced to one equation

\begin{eqnarray}\label{Eq29}
\frac{1}{\Lambda} \textrm{tanh}\left[\frac{{\beta}}{2}\Lambda\right]=\frac{1}{\Theta k_0},
\end{eqnarray}
where $\Lambda\equiv\sqrt{(\Theta k_0S_z)^2+4\lambda^2}$.

Noteworthy, Eq.~\eqref{Eq29} is completely symmetric in respect of collective spin $S_z$ and photonic field amplitude $\lambda$. Thereby, we can fix one of the order parameters (say, $\lambda$) and examine the phase transition properties for another one ($S_z$) by solving~\eqref{Eq29}.

The normal (non-SR) PM state with $S_z=0$, $\lambda=0$ characterizes some disordered phase for~\eqref{Eq29} formed upon condition $\Lambda=0$.
Concerning non-trivial solutions of Eq.~\eqref{Eq29} we are interested in the transition from $\Lambda=0$ to some spin ordering state with $\Lambda\neq0$ accompanied by the SR and/or PM-FM second order phase transitions, which are inherent to the Dicke and Ising models separately.

To be more specific, let us examine a phase boundary equation that describes the PM-FM phase transition occurring in the presence of the SR state ($\lambda=\lambda_c$). This equation can be easily obtained from~\eqref{H0_Sa} for vanishing but finite $S_z\to0$ and looks like
\begin{eqnarray}\label{PhB}
\frac{2\lambda_c}{\Theta k_0}=\textrm{tanh}[\lambda_c/T_{c\lambda}^{(2)}]. 
\end{eqnarray}
From~\eqref{PhB} for critical temperature $T_{c\lambda}^{(2)}$ we immediately obtain
\begin{eqnarray}\label{Tc_con30}
T_{c\lambda}^{(2)}=\frac{\lambda_c}{\textrm{tanh}^{-1}[\frac{2\lambda_c}{{\Theta}{k_0}}]}.
\end{eqnarray}

Critical temperature $T_{c\lambda}^{(2)}$ depends on network degree $k_0$, parameter $\Theta$, and another order parameter $\lambda_c$. In particular, it follows from~\eqref{PhB} that for $\lambda_c>\Theta k_0/2$ no phase transition occurs in the network system at any temperature. For $T<T_{c\lambda}^{(2)}$ and $\lambda<\lambda_c$ the FM SR phase represents a stable solution for the regular network system.

On the other hand, from~\eqref{Eq29} we can recognize critical temperature $T_{c,S}^{(2)}$ of the SR phase transition
\begin{eqnarray}\label{PhB2}
T_{c,S}^{(2)}=\frac{\Theta k_0 S_{z,c}}{2\textrm{tanh}^{-1}[S_{z,c}]}, 
\end{eqnarray}
setting in~\eqref{Eq29} $\lambda\simeq0$ for $S_z=S_{z,c}$.

In the vicinity of the PM non-SR state, assuming in~\eqref{PhB}-\eqref{PhB2} $\lambda_c\to0$ and $S_{z,c}\to0$, respectively, we can use in~\eqref{Eq29} approximated formula $\textrm{tanh}[\frac{\beta}{2}\Lambda]\approx \frac{\beta}{2}\Lambda $. In this limit critical temperatures~\eqref{Tc_con30},~\eqref{PhB2} are equal to each other $T_{c\lambda}^{(2)}=T_{c,S}^{(2)}=T_c^{(2)}$ and tend to
\begin{eqnarray}\label{Eq13}
T_c^{(2)}=\frac{1}{2}\ \Theta k_{0,c}.
\end{eqnarray}

Here, we consider the situation when the critical temperature in~\eqref{Eq13} implies the existence of critical degree $k_{0,c}$. The phase transition occurs in the networks possessing degree $k_0<k_{0,c}$.

Remarkably, the critical temperature in~\eqref{Eq13} tends to infinity in the thermodynamic limit at $N\to\infty$ for the complete regular graph with $k_{0}=N-1$. However, in practice  $T_c^{(2)}$ may be high enough but finite due to finite size effects which determined by number of nodes $N$.

Fig.~3 demonstrates a numerical solution of Eq.~\eqref{Eq29} established in 3D space by using $S_z$ and $\lambda$ variables. We examine here the particular case $\Theta k_{0}=\Omega_a=1$ for~\eqref{epsi},~\eqref{Eq29}. The SR FM domain appears within temperature window $0\leq T<T_c^{(2)}$ as a result of intersection of $\sigma$-plane ($z=\frac{1}{\Theta k_0}=1$) with a surface representing function $F(S_z,\lambda)=\frac{1}{\Lambda} \textrm{tanh}\left[\frac{{\beta}}{2}\Lambda\right]$, which is relevant to the left side of~\eqref{Eq29}. Clearly, the size of the domain is maximal in the zero temperature limit, see the black solid curve $AB$ in Fig.~3. This domain is characterized by equation $\sqrt{(\Theta k_0S_{z})^2+4\lambda^2}=\Theta k_0$ that may be obtained from  (28). Obviously, the domain  reduces with the temperature increasing up to $T_c^{(2)}=1/2$, see the red curves in Fig.~3.

\begin{figure}[h!]
\center{\includegraphics[width=0.8\linewidth]{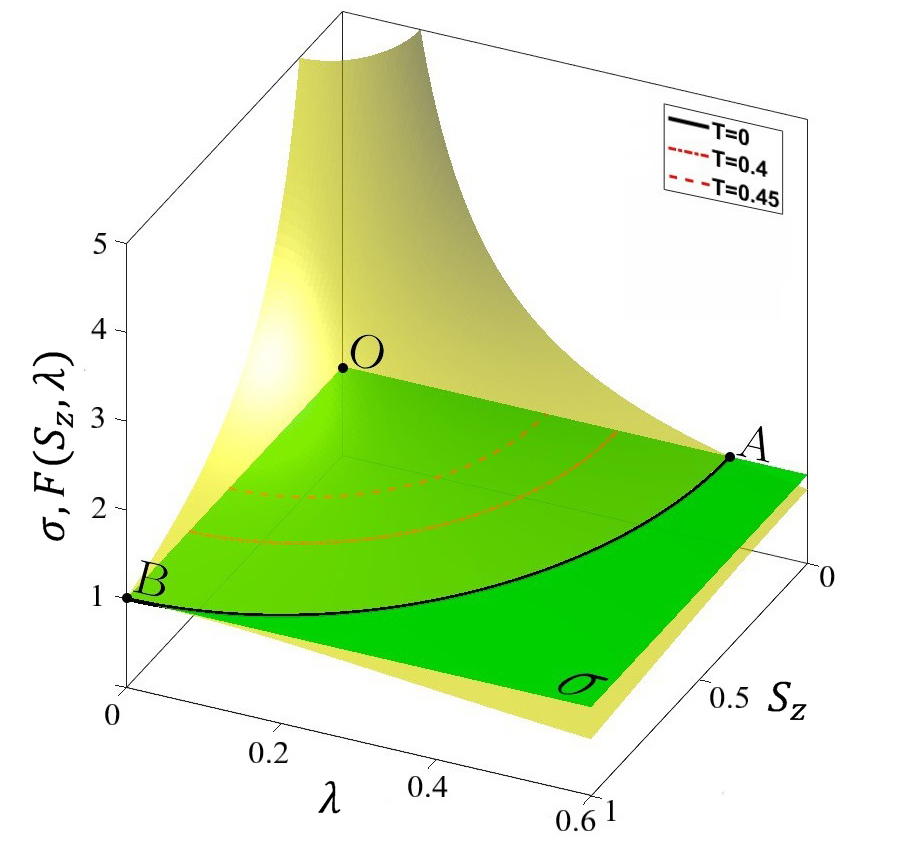}}
\caption {Function $F(S_z,\lambda)$ (yellow surface), $\sigma$-plane (green plane) vs. $S_z$ and $\lambda$ for $\Theta k_{0}=\Omega_a=1$ and temperature $T=0$.  The point $O$ indicates the PM normal (non-SR) state. The lines $OA$ and $OB$ correspond to the SR PM and normal FM phases, respectively, which occur within the suitable temperature domain. The red  curves indicate  the solutions of (28) at two different temperatures ($F(S_z,\lambda)$ is not shown for them).}
\label{1}
\end{figure}

The results obtained in Eqs.~\eqref{Tc_con30}-\eqref{Eq13} admit a simple physical interpretation given in Fig.~3. 

The point $O$ in Fig.~3  with coordinates $S_z=\lambda=0$ characterizes the disordered (PM non-SR) state $\Lambda=0$ being a solution of Eq.~\eqref{Eq29} at temperature $T_c^{(2)}$, cf.~\eqref{Eq13}.

The line $OB$ represents the phase  boundary for the SR phase transition in the presence of some FM state relevant to non-zero $S_{z,c}$ and characterized by critical temperature $T_{c,S}^{(2)}$ defined in~\eqref{PhB2}.

Remarkably, the line $OA$ represents the PM-FM phase   boundary in the presence of superradiance. Various values of $\lambda_c$ on $OA$ are inherent to critical temperature $T_{c\lambda}^{(2)}$. 
For example, the point $A$ corresponds to the phase transition at zero critical temperature that implies  $\lambda_c=\Theta k_0/2$, cf. (29).
 By means of definition (4), for $N\gg1$, we can represent critical photon number $N_{ph,c}$ required to achieve the phase transition for the complete graph as 
 
\begin{equation}\label{eq8}
N_{ph,c}\simeq\frac{\Theta^2 N^3}{4}.
\end{equation}

In some  applications of the SR phase transition in photonics two limiting cases are usually considered.  First, we can speak about convenient  lasing phenomena when the number of photons is much larger than the number of particles (nodes in our case), cf. ~\cite{Eastham, Chestnov}. Another limit that implies fulfillment of inequality   $N_{ph}{\ll}N$ is typically considered in the framework of polariton lasers occurring in the  strong matter-field coupling regime [36, 42, 43]. 
These  limits appear in experiments with exciton polariton BEC as two thresholds to the lasing effects, which possess different physical background  \cite{Yamamoto, SnokeL}. 
For the network system inequality $N_{ph,c}{\ll}N$ applied to the  critical photon number $N_{ph,c}$ together with (33)  implies that dimensionless spin-spin interaction strength $\Theta$ obeys the condition  $\Theta\ll2/N$. Since number of nodes $N$ is huge for many practical applications~\cite{Bianconi3,Suchecki},  $\Theta$ should be  small enough in this limit. 

\section{Phase transitions in complex networks}
\subsection{Superradiant phase transition in the random and scale-free networks}
\paragraph{Phase transitions in $H\to0$ limit.}
Now let us consider the phase transitions in the random and scale-free networks by means of Eqs.~\eqref{S}. 
Unfortunately, due to large set of parameters occurring in the Dicke-Ising model, it is hard to examine~\eqref{S} analytically in a general case. Here, we represent some important limiting cases, which admit simple treatments for the critical parameters obtained under the phase transition condition.
In particular,~\eqref{S} for $H=0$ yields
\begin{subequations}\label{S1}
\begin{align}
\begin{split}
F_1(S_z, \lambda)\equiv\int\limits^{k_{max}}_{k_{min}}{p(k)\frac{\Theta k^2}{\Gamma\langle{k}\rangle} \textrm{tanh}\left[\frac{{\beta}}{2}\Gamma\right]dk}=1,
\end{split} \label{S1a}\\
\begin{split}
F_2(S_z, \lambda)\equiv\frac{1}{\Omega_a} \int\limits^{k_{max}}_{k_{min}}{p(k)\frac{\textrm{tanh}[\frac{{\beta}}{2}\Gamma]}{\Gamma}}dk=1,
\end{split} \label{l1b}
\end{align}
\end{subequations}
where $\Gamma\equiv\sqrt{({\Theta}S_z k)^2+4{\lambda}^2}$.

\begin{figure*}[ht]
\begin{minipage}[h]{0.33\linewidth}
\center{\includegraphics[width=\linewidth]{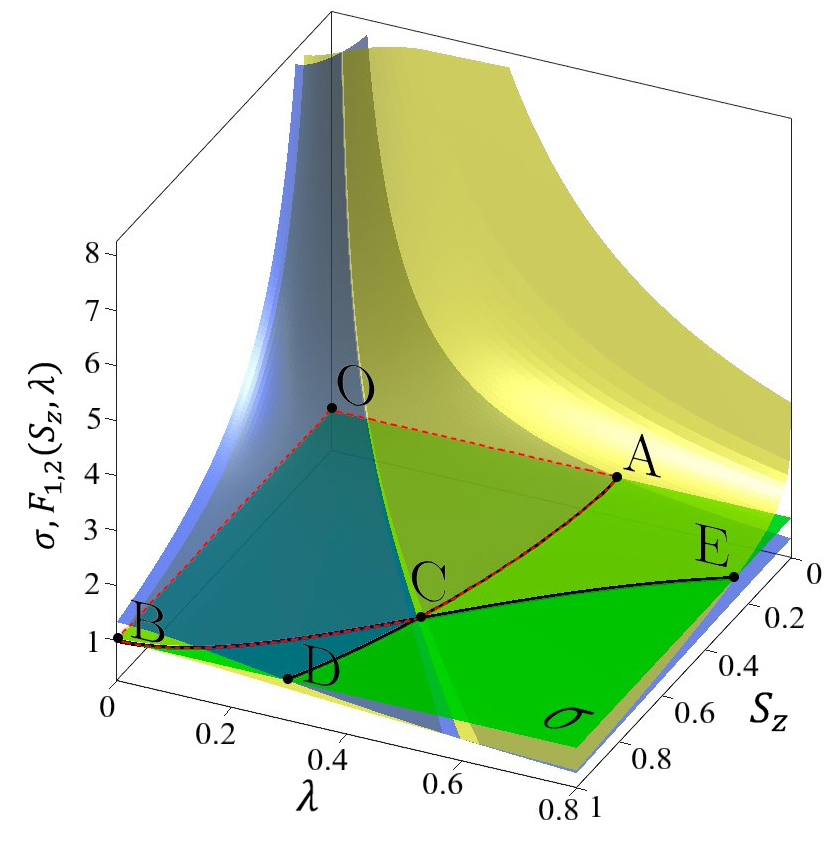} \\ (a)}
\end{minipage}
\hfill
\begin{minipage}[h]{0.65\linewidth}
\center{\includegraphics[width=\linewidth]{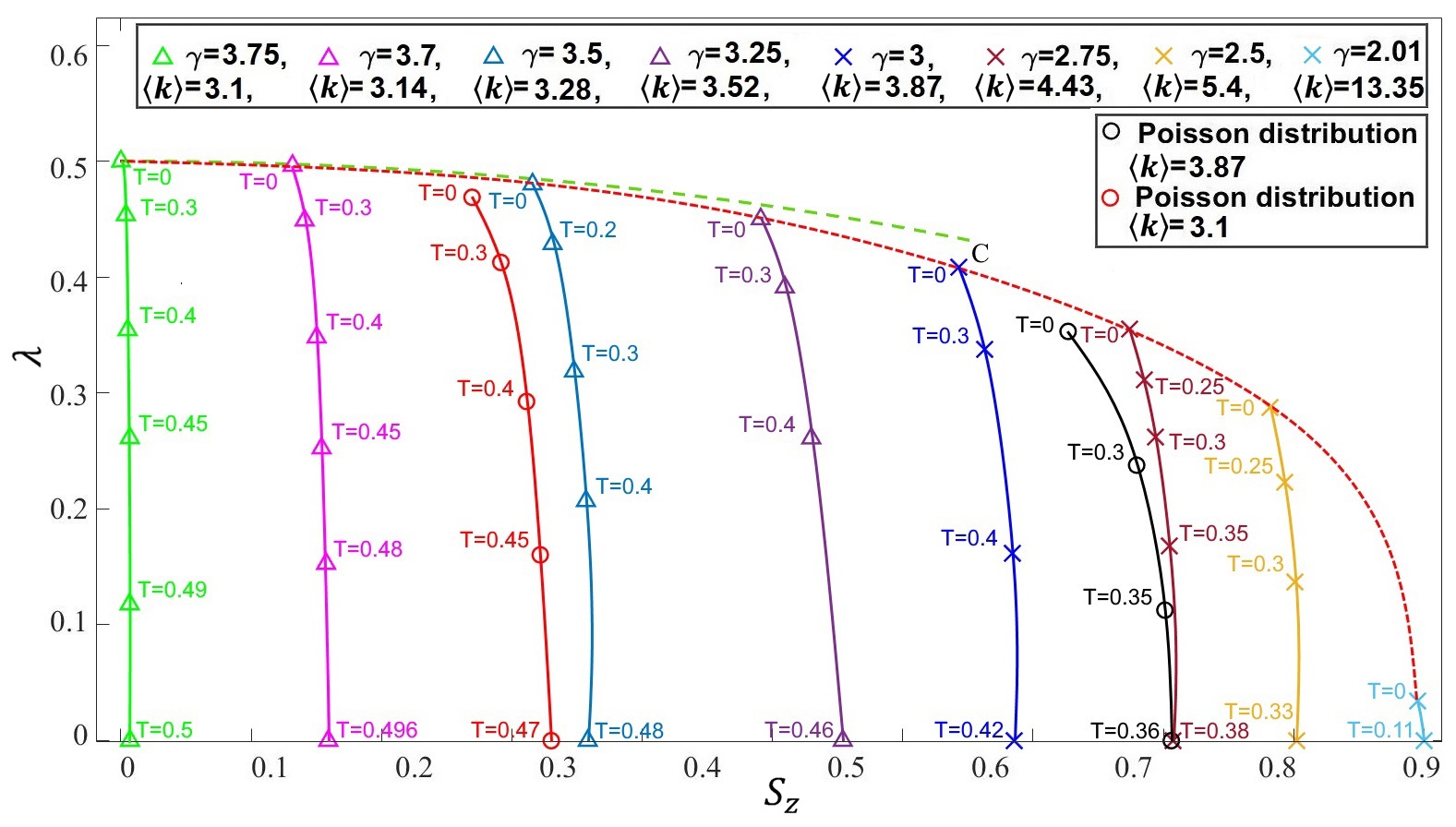} \\ (b)}
\end{minipage}
\caption{
(a)  Functionals $F_1$ (yellow surface), $F_2$ (blue surface) and $\sigma$-plane (green plane) vs. $S_z$ and $\lambda$ at temperature $T=0$ and for $\gamma=3$. (b) Phase boundaries for the crossing points in the $\lambda$ - $S_z$ plane. The parameters are: $\Theta=0.25$, $\Omega_a=1$, and $N=1000$. The maximal (minimal) degree for the scale-free networks   are $k_{max}=k_{min}N^{\frac{1}{\gamma-1}}$ 
(${k_{min}}=2$). For the random networks $k_{max}=11$ (for $\langle{k}\rangle=3.87$) and $k_{max}=9$ (for  $\langle{k}\rangle=3.1$) with ${k_{min}}=0$ are used, respectively. The green dashed line represent solution of Eqs. (35). See more details in the text.}
\label{ris:image1}
\end{figure*}

In a general case, critical temperature $T_c$ of the SR phase transition for the complex networks essentially depends on on particular degree distribution $p(k)$. The physical explanation of this fact looks as following.

The random and especially scale-free networks considered here support a locally (node) dependent interaction between the spins because of some specific topological features (hubs, clusters, etc). This interaction leads to the existence of  effective  local field (see~\eqref{eff_mag}) responsible for the establishment of some FM ordering in $z$-direction even without external magnetic field $H$. Thus, the SR phase transition appears as a result of the interplay between spin ordering in $z$- and $x$-directions and depends on topological features of the  network in $\lambda\to0$ limit.

In Fig.~4 we examine solutions of Eqs.~\eqref{S1} numerically. In particular, Fig.~4(a) displays the solution of Eqs.~\eqref{S1} for the BA network. We take the parameters $\Omega_a=1$, $\Theta {\langle}k{\rangle}=1$, which are maximally close to the ones for Fig.~3. Graphically, this solution looks as the crossing point $C$ for $\sigma$-plane  and functionals $F_1(S_z,\lambda)$ (yellow surface) and $F_2(S_z,\lambda)$ (blue  surface), which represent the left-hand sides of Eqs.~\eqref{S1a} and~\eqref{l1b}, respectively, cf. Fig.~3. The intersection lines $AD$ and $BE$ confine the domains of the FM and/or SR phases, which correspond to the solutions of Eqs.~\eqref{S1a} and~\eqref{l1b} separately. The dashed red line restricts the area where the FM and SR phases coexist. This area is maximal at zero temperature.

Thus, the only one crossing point $C$ in Fig.~4(a) separates different phases, while in Fig.~3 this role is performed by the curve, which implies the phase boundaries in the $OA$ and $OB$ directions.

Fig.~4(b) establishes the numerical solutions of Eqs.~\eqref{S1} for various temperatures and values of $\gamma$ for the scale-free (crosses, triangles) and random (circles) networks. The solution of Eqs. (34)  for  scale-free networks exists within the  domain $2<\gamma \leq3.75$, where $\langle{k}\rangle$ possesses moderate values, see Fig.~2. For example, consider the dark blue curve  in Fig.~4(b) (marked by crosses) that corresponds to the BA network with $\gamma=3$. The curve starts at $T=0$ in the crossing point $C$, which is the same point as in Fig.~4(a), and then moves toward  $S_z$ axis with the temperature increasing. Critical temperature $T_c$ of the phase transition to the superradiance is obtained at the point where $\lambda=0$; it is equal to $T_c=0.42$. The functional in~\eqref{S1a} for this point is $F_1(S_z,0)=\int\limits^{k_{max}}_{k_{min}}{p(k)\frac{k}{S_z\langle{k}\rangle} \textrm{tanh}\left[\frac{{\beta_c}}{2}\Theta k S_z\right]dk}$ and corresponds to the FM state with $S_z\simeq 0.68$, see Fig.~4(b). This feature of $F_1(S_z,0)$ manifests the SR phase transition in the presence of the FM state that occurs at zero external field $H$, cf.~Fig.~3.

The dashed red line in Fig.~4(b) indicates the fact that $T_c$ diminishes for the scale-free network with $\gamma$ reduction. Simultaneously, the value of magnetization $S_z$ increases. This is not surprising, since without the external classical field, $H=0$, the network spin system aspires to increase the ordering state at lower temperatures. In particular, Eqs.~\eqref{S1} admit of the point in the $\lambda$ - $S_z$ plane (marked by bright blue crosses in Fig.~4(b)) that corresponds to the ordered collective spin state ($S_z\simeq1$) obtained at critical temperature $T_c\simeq0.11$ for $\gamma=2.01$. 

Remarkably, the same dashed (red) curve in  Fig.4(b) exhibits vanishing of  the order parameter $\lambda$ with increasing average degree  $\langle{k}\rangle$. We can elucidate this feature of the Dicke-Ising  model  examining the asymptotic solution of (34) in zero temperature limit. We assume in  (34) that $\Gamma\simeq \overline\Gamma\equiv\sqrt{({\Theta} S_z \langle k\rangle)^2+4{\lambda}^2}$, supposing that approach  $k\approx\langle k\rangle$ is valid. In this limiting case from (34) we obtain
\begin{subequations}\label{S11}
\begin{align}
\begin{split}
\Omega_{a} \Theta \zeta=1,
\end{split} \label{S11a}\\
\begin{split}
\Omega_a \overline\Gamma=1.
\end{split} \label{111b}
\end{align}
\end{subequations}

The green dashed curve  in Fig.4(b)  represents the solution of Eqs.(35) for $\langle{k}\rangle=3.1$ ($\zeta\simeq4$). From Fig.4(b) it is clearly seen that 
the approach we use here is valid within degree exponent $3.75\leq\gamma \leq3.25$, where the green dashed line approaches the red one, which is relevant to numerical solutions of (34). For  $\gamma<3.25$ the discrepancy between two dashed curves grows and  self-consistent solutions of Eq. (35) do not exist.

The phase boundaries  for the random networks are represented  in Fig.~4(b) by two (red  and black coloured) curves with circles. To find the upper natural cutoff $k_{max}$ we explore arguments represented in Sec.~III. The difference between the curves is determined by the value of $\langle{k}\rangle$ and exhibits general tendency of vanishing $\lambda$ with increasing $\langle{k}\rangle$.

Now let us analytically examine Eqs.~\eqref{S1}  when they admit  some simplifications. Remarkably, Eqs. ~\eqref{S1} possess  simple treatments for the PM-FM phase transition boundary that occurs in the complex networks at some SR state, for given $\lambda_c$ and  vanishing $S_z$. 
In particular, the green curve with triangles in Fig.~4(b) is relevant to this limit. 
Solving Eqs.~\eqref{S1} in the limit of vanishing $S_z$ for the critical temperature we get 
\begin{eqnarray}\label{Tc_con}
T_c=\frac{\lambda_c}{\textrm{tanh}^{-1}[\frac{2\lambda_c}{{\Theta}{\zeta}}]}
\end{eqnarray}
with condition (35a) that accounts for the solution of~(34b) in the same limit, cf.~(27).

Eq.~\eqref{Tc_con} represents a generalization of~ (30) obtained for the regular networks. The topological properties of the network in ~\eqref{Tc_con}  are taken into account at the macroscopic level using the $\zeta$- parameter.

In the vicinity of the disordered state at $S_z\sim0$, $\lambda\sim0$ Eq.~\eqref{S1} simplifies; for the critical temperature of the transition to the ordered state we can obtain
\begin{eqnarray}\label{Eq13R}
T_c=\frac{1}{2} \Theta \zeta.
\end{eqnarray}

Eq.~\eqref{Eq13R} represents a generalization of Eq.~\eqref{Eq13} for the complex networks in the Dicke-Ising model framework; in~\eqref{Eq13R} and thereafter we omit the upper indices for the critical temperature. It is noteworthy that the critical temperature (37) obtained for Dicke-Ising model  coincides with that for the Ising  model on networks, cf. [14, 19, 22, 24]. Moreover, this critical temperature agrees with results obtained rigorously by other approaches, cf. [1, 18].

From ~\eqref{Tc_con}, ~\eqref{Eq13R} it follows that since $\zeta\to\infty$ for the scale-free network in the thermodynamic limit, the critical temperature of the PM-FM phase transition is also infinite. However, the finite critical temperature  for the phase transition occurs in ~\eqref{Eq13R} due to the finite size effects for the scale-free network, see TABLE I. 

In Fig.~5 we plot dependence~\eqref{Tc_con}
for the critical temperature $T_c$ of PM-FM phase transition versus order parameter $\lambda$ for the scale-free (green curves) and random (the red dashed curve) networks, respectively. The  parameters for the scale-free network in Fig.~4 are relevant to the ones established for the green  curve with triangles in Fig.~4(b). The phase boundaries shown in Fig.~5 separate the PM state (it is shown as a shadow region for the scale-free network) and the FM SR state. 
Notably, ~\eqref{Eq13R} may be obtained from~\eqref{Tc_con}  in the limit of $\lambda_c\to 0 $.  In particular, the crossing point of the solid green curve in Fig.~5 with line $\lambda=0$ determines the critical temperature established by Eq. \eqref{Eq13R} for the  scale-free network. In this limit PM-FM phase transition leads to establishment of some ordering  FM non-SR state.

It is instructive to compare Eqs. \eqref{Tc_con}, \eqref{Eq13R} with the results obtained in the framework of other mean-field theories suitable for the Ising model, see \cite{Napoli2, Bianconi2, Dorogovtsev}. 
In  ~\cite{Napoli2} authors consider the approximation of   the Ising model on a BA network  by the  effective long-range homogeneous Ising model  providing effective spin-spin interaction  strength determined through the average  degree of the BA network.  The result for the effective long-range Ising model may be obtained from \eqref{Eq13R}  setting $\zeta\simeq \langle k\rangle$ that implies $\langle k^2\rangle \approx \langle k\rangle^2$. 
Physically, such a condition means ``homogenization" of the scale-free network, which may be relevant for large  degree exponent $\gamma$, cf. \cite{Dorogovtsev}.  As it follows from TABLE I, such an approximation error for the effective long-range Ising model grows with the number of nodes $N$ as $\zeta\ / \langle k\rangle = \frac{1}{4} \textrm{ln}(N)$. Moreover, as it follows from Fig. 2, discrepancy between  $\zeta$ and $\langle k\rangle$ is large enough  within domain   $1<\gamma<3$, where the  effective long-range Ising model seems to be inapplicable. 

To be more specific, in Fig.5  we examine the effective long-range Ising model for the scale-free network with degree exponent $\gamma=3.75$; the dashed green curve in Fig.5 represents the phase boundary in this case. This mean-field approach accuracy may be estimated from TABLE I; it is determined by ratio  $\zeta\ / \langle k\rangle\simeq1.5(1-N^{-3/11})\approx1.26$ with $N=1000$.   
In this approximation  the critical temperature (36) for the scale-free network behaves as 
$T_c=\lambda_c/\textrm{tanh}^{-1}[\frac{2\lambda_c}{{\Theta}{\langle k\rangle}}]$, which is reminiscent to the regular network with $\langle k\rangle=k_0$, cf. (30) and   ~\cite{Napoli2}. Thereby, some discrepancies for the green curves in Fig.5 may be explained due to the scale-free network ``homogenezation" procedure.

Red curves in Fig.5  establish temperature dependence for random networks.  The dashed curve represents phase boundary in accordance with Eq.~\eqref{Tc_con}, i.e. in the limit of $S_z=0$.  However, as it follows from numerical simulations given  in Fig.4(b) we can  treat collective spin   $S_z$ in (34) as some  constant possessing average value   within  temperature  domain being under consideration  if the variation of $S_z$ with temperature is  not so large. 

In Fig.5 we represent the solid red curve for the random network possessing  $S_z\approx0.3$ and relevant to the red curve  with circles in Fig.4(b). Within this limit the temperature dependence on  $\lambda$ is obtained from (34a) and approaches 
\begin{eqnarray}\label{T}
T=\frac{\overline\Gamma}{2\textrm{tanh}^{-1}[\frac{\overline\Gamma}{{\Theta}{\zeta}}]}.
\end{eqnarray}
 
As clearly seen from Fig.5, for moderate values of collective spin $S_z$ and coupling strength $\Theta$  the temperature dependence (the dashed curve) approaches its phase boundary. At the same time, dependence for the random networks is close to the solid green curve that characterizes the scale-free network  in  the random domain, cf. \cite{Barabook}.

\begin{figure}[h]
\center{\includegraphics[width=1\linewidth]{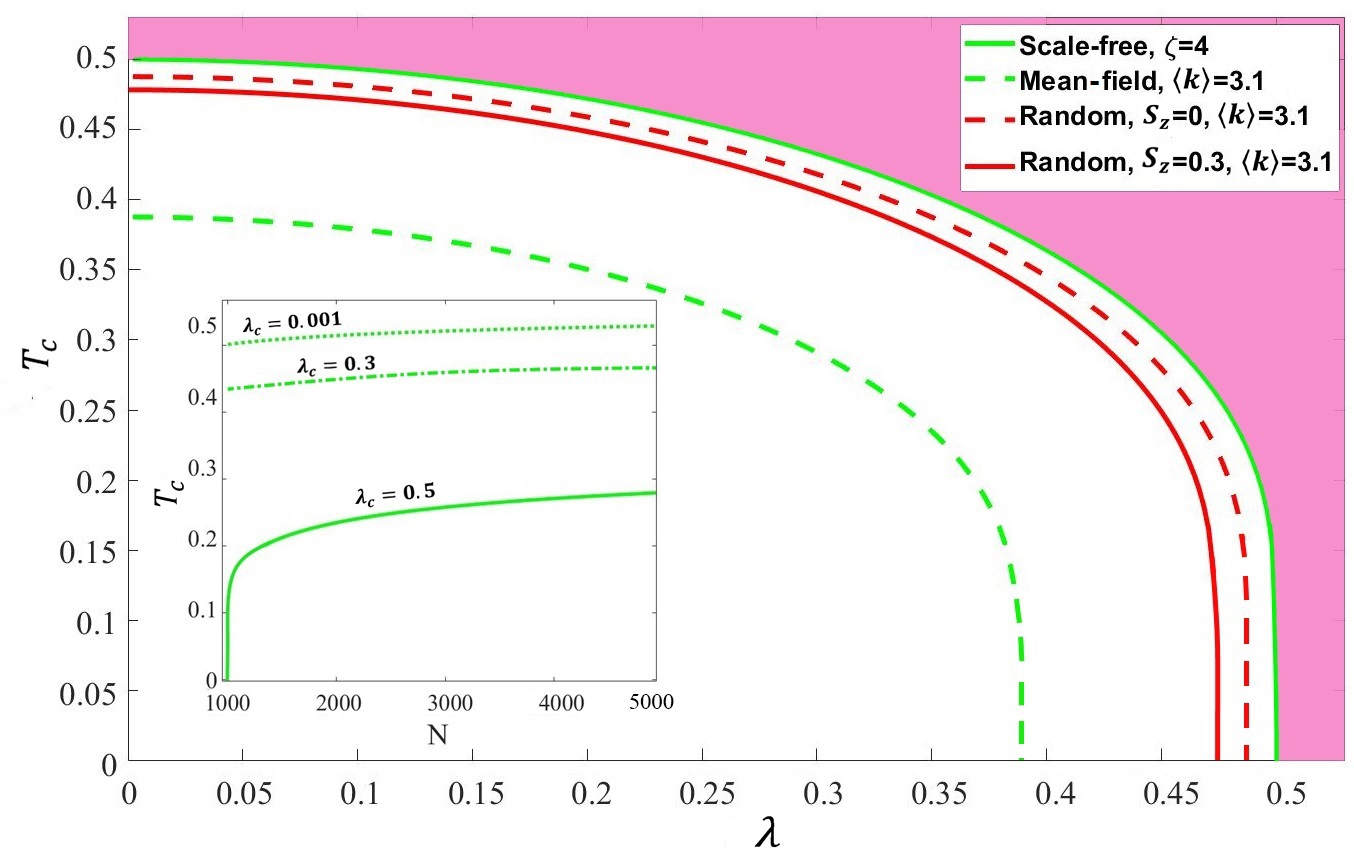}}
\caption{Dependence of  temperature $T$ versus order parameter $\lambda$ for the scale-free network (green curves) with $\gamma=3.75$, $S_z\simeq0$  and random one  with $S_z\simeq0.3$ (solid red curve), $S_z\simeq0$ (dashed red curve), respectively. 
The parameters are: $\Theta=0.25$, $\Omega_a=1$, $N=1000$, and ${{\langle}k{\rangle}}=3.1$, cf. Fig.4(b). The shadow region corresponds to the PM phase $S_z=0$. The inset indicates the dependence of critical temperature $T_c$ on number of nodes $N$ calculated for the same scale-free network at different $\lambda_c$.} 

\label{1}
\end{figure}

The finite size effects may be obtained from the  last column in TABLE I that displays the critical number of nodes, which is required to obtain  phase transition for a given network system.
As example, from (36) for the BA scale-free network that imposes $\gamma=3$ the critical temperature of the phase transition is $T_c=\frac{1}{8}\ \Theta {{\langle}k\rangle} \textrm{ln}(N_c)$ that immediately defines the critical number of nodes 
\begin{eqnarray}\label{Eq14R}
N_c=e^{8T_c/\Theta {{\langle}k\rangle}}.
\end{eqnarray}

Equation~\eqref{Eq14R} implies critical clustering coefficient ${\langle}C_c\rangle\sim (\textrm{ln}(N_c))^2/N_c$ permissible for the phase transition. 

The inset in Fig.~5 exhibits  the behaviour of critical temperature $T_c$  on number of nodes $N$ as it follows from ~\eqref{Tc_con}.  Since the number of nodes $N$ is large enough, we restrict ourselves to region $N\geq1000$. The curves in Fig.~5 inset constrain the region of superradiance.
In particular, these curves  demonstrate that critical temperature $T_c$ grows  starting from some certain value that corresponds to the critical number of  nodes where the solution of~\eqref{Tc_con} exists.

\begin{figure*}[ht]
\begin{minipage}[h]{0.49\linewidth}
\center{\includegraphics[width=\linewidth]{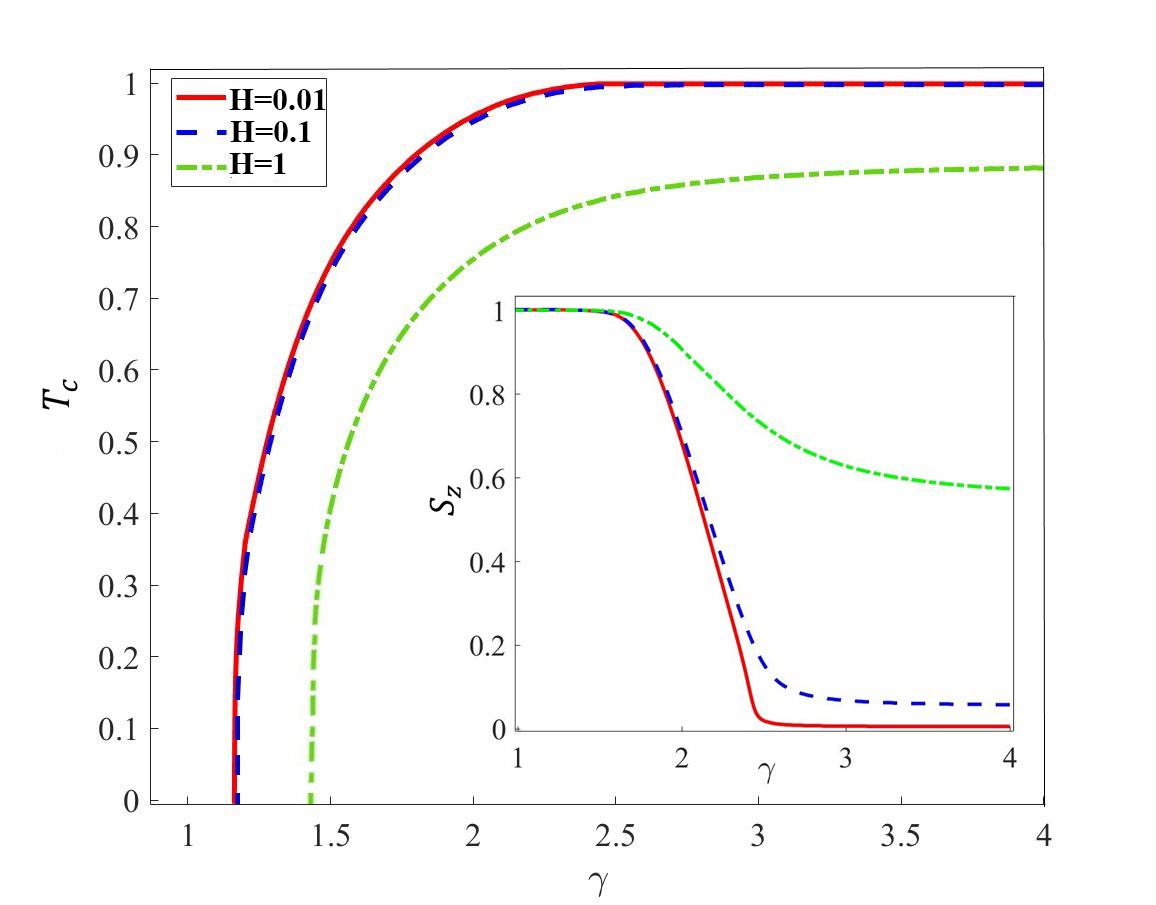} \\ (a)}
\end{minipage}
\hfill
\begin{minipage}[h]{0.49\linewidth}
\center{\includegraphics[width=\linewidth]{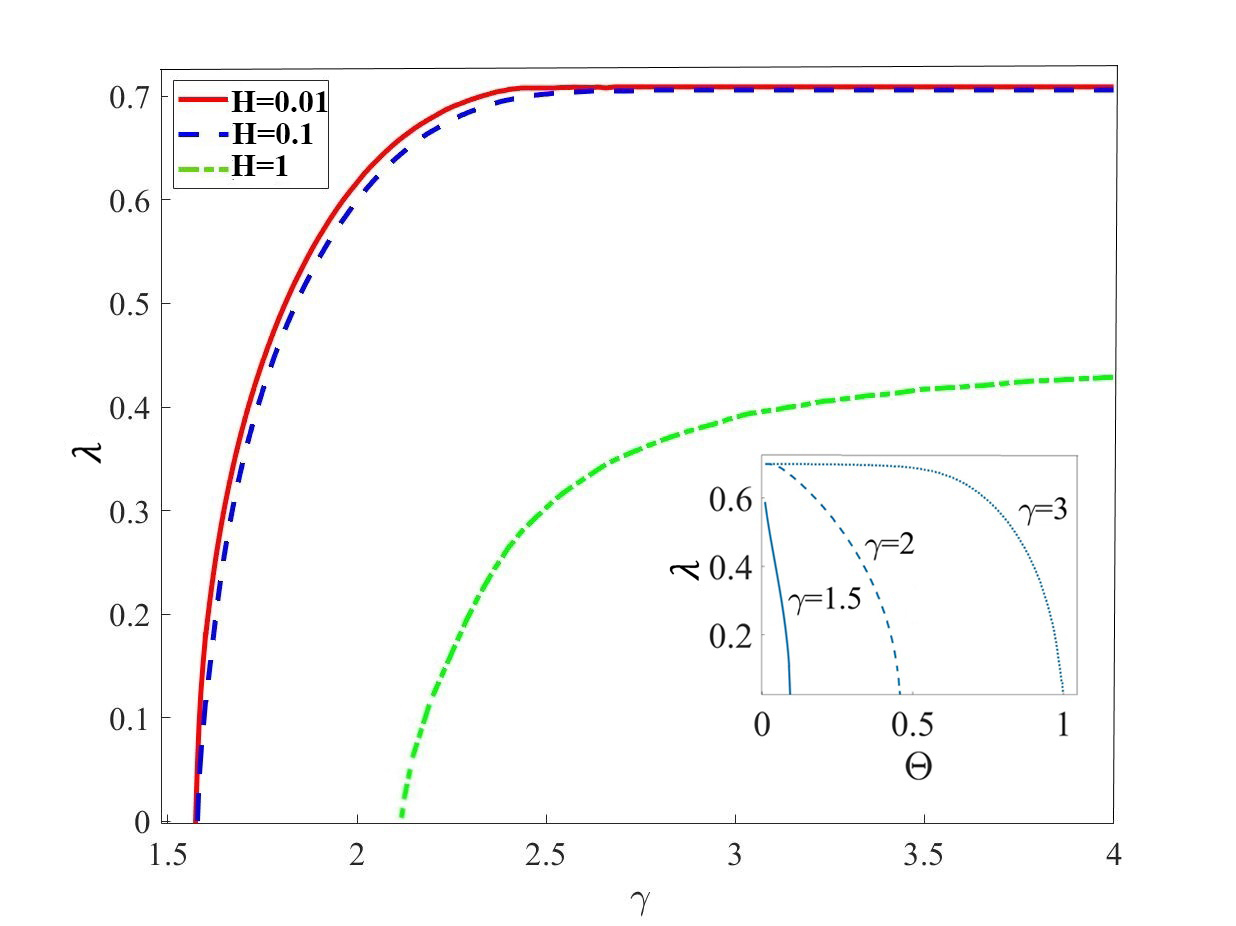} \\ (b)}
\end{minipage}
\caption{Dependence of (a) critical temperature $T_c$ and (b) order parameter $\lambda$ vs. $\gamma$. The parameters are: $\Theta=0.15$, $N=1000$, $\Omega_a=0.5$, ${k_{min}}=1$, $k_{max}=k_{min}N^{\frac{1}{\gamma-1}}$  for (a) - $\lambda=0$, and  for (b) - $T=0.8$, respectively; ${\langle}k{\rangle}$ specified in TABLE~I. The inset in (a) exhibits total spin component $S_z$ vs. $\gamma$ for the dependence given in (b) at $T=0.8$. The inset in (b) demonstrates the dependence of $\lambda$ on $\Theta$ at $H=0.1$ and $T=0.8$ but for different values of $\gamma$: $\gamma=1.5$, $\gamma=2$, and $\gamma=3$.}
\label{ris:image1}
\end{figure*}

\paragraph{Phase transitions at non-vanishing classical field $H$ and for scale-free networks.} 
The aim of this part is to study the phase transition problem for the scale-free network within a large domain of power degree $\gamma$ and non-vanishing classical field $H$. We now search for non-trivial solutions $\lambda\neq0$ and $S_z\neq0$ of Eqs.~\eqref{S}.

In Fig.~6 we represent numerical solutions of Eqs.~\eqref{S} for critical temperature $T_c$ and order parameter $\lambda$ as functions of $\gamma$. 

Notice, collective spin component $S_z$, as it follows from Fig.~6(a), approaches the FM state within the domain $1<\gamma<1.7$ of the anomalous regime for the scale-free networks. Roughly speaking, the spin system exhibits the fully ordered state. Fig.~6(b) shows that the influence of another order parameter $\lambda$ is not so important in the anomalous regime where $S_z\simeq1$. 
In fact, in this case we deal with the situation valid for the familiar Ising model without the transverse field. Moreover, the thermal fluctuations have no ability to break the FM state at any finite temperatures within the domain $1<\gamma<3$, cf.~\cite{Dorogov1}. 

The behaviour of the spin system within $1<\gamma<1.7$ domain completely depends on average degree ${\langle}k{\rangle}$ that grows rapidly, as it follows from Fig.~2. In particular, in the anomalous regime the size of the largest hub $k_{max}\propto N^{\frac{1}{\gamma-1}}>N$, and the number of links connected to the largest hub increases faster than the size of the network, $N$. We expect to obtain a strong spin-spin interaction within this domain. Thus, we assume that randomly chosen $k$ possesses large values due to the power law degree $p(k)$, see Fig.~1(d). In this case, for large $k$ we can use approximation $\textrm{tanh}(\frac{{\beta}}{2}({\Theta}S_zk+H))\approx1$ in Eq.~\eqref{Sa}, which allows to obtain $S_z\simeq1$ for the collective spin, see the inset in Fig.~6(a).

Remarkably, in the anomalous regime the main contribution to effective magnetic field $H_{eff}$ comes from the term that characterizes the spin-spin interaction and depends on network parameter ${\langle}k{\rangle}$, see~\eqref{eff_mag},~\eqref{eff_mag2}. From Fig.~6 we can see that behaviour of the spin system for small values of external field $H$ (the red and blue curves) are practically still the same. 

In the presence of the interaction with the quantized transverse field, $S_z$ abruptly vanishes at small values of classical magnetic field $H$, which is clearly seen from the inset in Fig.~6(a). In Fig.~4(b), we have already specified this behavior of $S_z$ within the scale-free and random  domains ($2<\gamma\leq3.75$), respectively.

In particular, we perform the analysis of Eqs.~\eqref{S} in the relatively high temperature limit for non-zero field $H$; the spontaneous magnetization can be represented in the following form (cf. (25))
\begin{eqnarray}\label{S2}
S_z=\frac{{\beta}H}{2-{\beta}{\Theta}\zeta},
\end{eqnarray}
that defines the FM state with vanishing $S_z$ in the presence of superradiance. $S_z$ approaches some constant value $S_z\approx\frac{{\beta}H}{2}$ for vanishing $\zeta$ that clearly exhibits the dependence of $S_z$ in the inset in Fig.~6(a) in the network system. 

This happens due to the critical temperature increasing and establishing the finite superradiant field amplitude $\lambda$ that characterizes the magnetization in $x$-direction; critical temperature $T_c$ in Fig.~6(a) approaches some non-zero value for $\gamma\geq 2.5$. In particular, the superradiant field promotes the spin flipping that occurs on the scale-free network nodes, see Fig.~6(b). In other words, the establishment of non-vanishing transverse field introduces some disordering in the collective spin component in $z$-direction. In contrast, classical magnetic field $H$ tends to preserve the ordered state of collective spin $S_z$, see the green curves in Fig.~6(a) which are plotted for value $H=1$. The absolute value of the established superradiant field amplitude $\lambda$ vanishes in this case, see the green curve in Fig.~6(b). 

The inset in Fig.~6(b) demonstrates the suppression of order parameter $\lambda$ with increasing of $\Theta$. Such a behaviour happens due to the contribution of the spin-spin interaction energy into total magnetization in $z$-direction. 
As it follows from Eqs.~\eqref{eff_mag},~\eqref{eff_mag2} the increasing of $\Theta$ for a given degree exponent $\gamma$ leads to  the increasing of effective field $H_{eff}$ that enhances magnetization in $z$-direction suppressing it in $x$-direction. For the degree exponent less than two, the domain where the SR state exists, $\lambda\neq1$, is narrow enough, see the solid curve with $\gamma=1.5$ in the inset in Fig.~6(b). As we discussed before, in this case the spin system possesses some ordered state, as it is shown in the inset in Fig.~6(a). At larger values of degree exponent $\gamma$, the SR state domain exists within a large $\Theta$-parameter window (that indicates inset in Fig.~6(b)) due to establishing of the non-vanishing superradiant field for $\gamma>2$, as it is shown in Fig.~6(b). 
 
\subsection{Quantum phase transitions}

The quantum phase transition is obtained in the zero-temperature limit setting in Eqs.~\eqref{S} $\beta\to \infty$ that presumes $\textrm{tanh}(\frac{{\beta}}{2}\Gamma)\simeq1$. From~\eqref{S} for small enough $\lambda$ in this case we obtain: 
\begin{subequations}\label{S-T0}
\begin{align}
\begin{split} 
S_z=1-\frac{2\lambda^2}{\langle{k}\rangle} \int\limits^{k_{max}}_{k_{min}}{\frac{kp(k)}{(\Theta S_zk+H)^2}} dk,
\end{split}\\
\begin{split}\label{OmegaT0}
{\Omega}_a=\int\limits^{k_{max}}_{k_{min}}{\frac{
p(k)}{(\Theta S_zk+H)}\left(1-\frac{2\lambda^2}{(\Theta S_zk+H)^2}\right)}dk.
\end{split}
\end{align}
\end{subequations}
As clearly seen from Eqs.~\eqref{S-T0}, the spin system is fully ordered, i.e. $S_z=1$ for $\lambda=0$. This situation agrees with the results obtained for 
Fig.~4(b) (the bright blue curve) 
and Fig.~6(a) at non-zero temperatures and relatively small $\gamma$.

The SR quantum phase transition, with $S_z=1$, occurs for the critical value of transverse field frequency ${\Omega}_{a,c}$ that is determined from~\eqref{OmegaT0} and looks like
 \begin{eqnarray}\label{Om_c}
{\Omega}_{a,c}=\int\limits^{k_{max}}_{k_{min}}{\frac{
p(k)}{\Theta k+H}}dk.
\end{eqnarray}

Eqs.~\eqref{OmegaT0},~\eqref{Om_c} result in a simple dependence for order parameter $\lambda$ 
\begin{eqnarray}\label{Orderpar}
{\lambda}=\lambda_0\sqrt{1-
\frac{\Omega_{a}}{\Omega_{a,c}}},
\end{eqnarray}
where  ${\lambda_0}=\sqrt{\frac{\Omega_{a,c}}{2I}}$ is the order parameter in the limit $\Omega_{a}\to 0$, $I\equiv{\int\limits^{k_{max}}_{k_{min}}{\frac{p(k)}{({\Theta}{S_z}k+H)^3}}dk}$.

We can simplify Eqs.~(42), \eqref{Orderpar} in the vicinity of $H=0$ for the FM state, $S_z=1$. In particular, for the regular network we obtain $\lambda_0=0.71\Theta k_0\equiv0.71\Theta {\langle{k}\rangle}$, cf. Fig.3. For the BA network  possessing very large number of nodes $N$ we get  $\lambda_0\simeq 0.46\Theta {\langle{k}\rangle}$, see  TABLE I,  $\gamma=3$.
Thereby, the approach used in Eq.~\eqref{Orderpar} is valid for the network parameters obeying condition $\Theta {\langle{k}\rangle}\ll1$.

On the other hand, for finite field $H$ close to the disordered  state with $S_z=0$ from Eq.~\eqref{Orderpar}, we can deduce that the quantum phase transition has no dependency on the network characteristics and results in the establishment of small transverse, $m_{x}$, and longitudinal, $m_{z}$, magnetizations proportional to ${\lambda_0}=H/\sqrt2\ll1$ and $\Omega_{a}/\Omega_{a,c}\ll1$, respectively. 

\begin{figure}[ht]
\center{\includegraphics[width=1\linewidth]{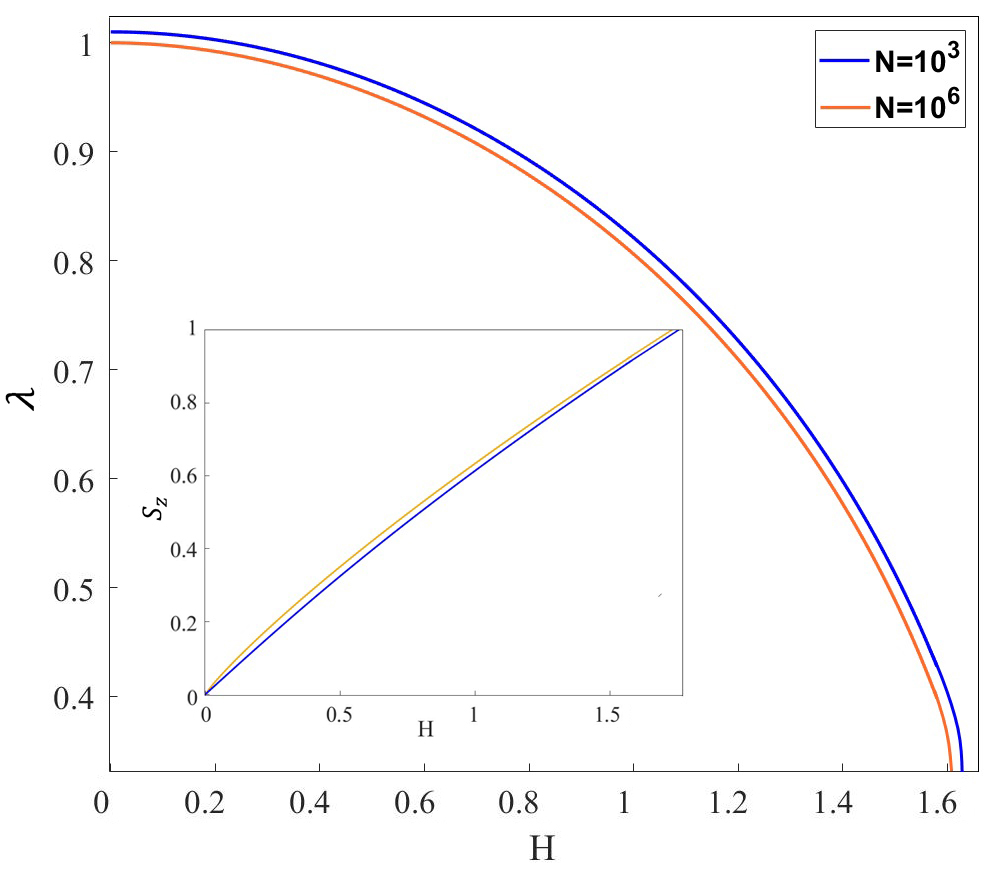}}
\caption{Order parameter $\lambda$ versus local field $H$ for the BA network, $\gamma=3$, at $T_c=0$  for different number of node $N$. The dependence of magnetization $S_z$ on $H$ is shown in the inset. Other parameters are: $\Omega_a=0.5$, $\Theta = 0.15$, $k_{min} = 1$, $k_{max}=k_{min}\sqrt{N}$.}
\label{1}
\end{figure}

In Fig.~7 we plot the dependence of the order parameters $\lambda$ and $S_z$ as functions of the magnetic field $H$, which represent the numerical solutions of Eqs.~\eqref{S} for the BA network with different number of nodes $N$ in the zero temperature limit. The SR quantum phase transition occurs for non-zero $H$ when the magnetization attained the FM state with $S_z=1$. In the vicinity of the phase transition point $H_c$, the order parameter $\lambda$ behaves as $\lambda \propto\sqrt{1-H/H_c}$, cf.~\eqref{Orderpar}. 

\section{Conclusion}

To summarize, we have considered the problem of the superradiant phase transition in the network structures. The Dicke-Ising model is developed to elucidate the second order phase transition for the regular, random, and scale-free networks. The model concerns the spin-1/2 (two-level) systems located in the network nodes and placed in the local classical magnetic and weak quantum transverse (photonic) fields. 
Applying the mean-field approach,  familiar in quantum optics, we have obtained the set of Eqs.~\eqref{S}, which describes two order parameters $S_z$ and $\lambda$ relevant to magnetization along $z$- and $x$-axes, respectively.  In particular, Eq.~\eqref{Sa} characterizes the collective weighted spin $z$-component, $S_z$. It is relevant to the PM-FM phase transition problem. Eq.~\eqref{lb} establishes the normalized transverse field amplitude, $\lambda$, which corresponds to the phase transition to the superradiance for the spin network system. The SR state with $\lambda\neq0$ is characterized by a non-zero temperature dependent energy gap that takes place for  the spin system. 

To be more specific, we have examined the annealed networks with the regular (fixed) degree number, the random network with the Poisson degree distribution, and the scale-free networks possessing the power-law degree distribution with degree exponent $\gamma$ that inherent to the region $1<\gamma\leq4$.

We have considered the problem of the phase transition for the Dicke-Ising model in two limits of classical field $H$. 
The superradiance occurs in the ferromagnetic spin system in the limit of the finite (non-zero) classical magnetic field. For the regular networks we have obtained analytically a simple equation~\eqref{orderp} analogous to the familiar law of temperature dependence for the gap systems (superconductors, ensemble of two-level systems interacting with quantized electromagnetic field) possessing the continuous second order phase transition. 
In this case, the transition to the SR state is followed by the establishment of the spontaneous magnetization with the quantum transverse field in $x$-direction. For the network that represents a complete graph, the critical number of nodes defines the necessary condition to attain the SR state. 

The physical picture becomes richer and more complicated for vanishing classical external field $H$. We have shown that for the regular networks Eqs.~\eqref{S} reduce to one (gap-like) equation~\eqref{Eq29}, which is completely symmetric in respect of order parameters $S_z$ and $\lambda$. Physically, we can recognize the phase transition occurring in the network system at the critical temperature as a crossover from the disordered state with $S_z=0$, $\lambda=0$ to some ordered state possessing $S_z\neq0$ (FM state) and/or $\lambda=0$ (SR state).

To achieve the ordering state for the regular networks possessing constant degree $k_0$, one can fix vanishing collective spin $S_z$ and then consider the SR phase transition across order parameter $\lambda$. Alternatively, we can fix $\lambda$ and then obtain the PM-FM phase transition.

It is important to note that in the $S_z\simeq0$ and $\lambda\simeq0$ limits the critical temperature of the disorder-order phase transition for examined  networks  exhibit the finite size effects.  Remarkably, our mean-field  approach  allows to account degree correlations in the scale-free network structure which we describe by the parameter $\zeta$, see (13). 
 In particular, the obtained critical temperature may be high enough but finite; it  depends on average degree ${\langle}k{\rangle}$ for the regular network and  statistical properties ($\zeta$-parameter) for the scale-free and random networks, respectively.
The analytical and numerical simulations which we performed for the scale-free networks within scale-free and random regimes, allow to conclude that the order parameter, $\lambda$, vanishes with increasing average degree  ${\langle}k{\rangle}$ and increasing collective spin component $S_z$.
The ``homogenezation" of the scale-free network structure, that presumes  $\zeta\approx {\langle}k{\rangle}$, may be relevant for large degree exponent $\gamma$.   
In this sense, our results agree with the ones obtained for the scale-free networks possessing the power degree distribution achieved by other methods, cf. \cite{Bianconi2, Dorogovtsev}. As seen from Fig.5 (the green curves), the mean-field approach based on scale-free network ``homogenezation" leads to some discrepancies within  critical temperature region, cf.  ~\cite{Napoli2}.

In general, the features of the complex networks phase transition strictly depend on power degree $\gamma$. In the anomalous regime, $1<\gamma<2$, the effective spin-spin interaction, which may be described in terms of effective magnetic field, see~\eqref{eff_mag},~\eqref{eff_mag2}, dominates due to large average degree ${\langle}k{\rangle}$. In particular, at low $\gamma$ the FM state with $S_z=1$ establishes; this state is very hard to alter by means of the spin interaction with the weak quantized field or even with the moderate external magnetic field. However, the situation changes with $\gamma$ increasing.  Diminishing  the effective spin-spin interaction leads to the suppression of collective $S_z$ spin component and creates an enabling environment for the PM-FM and SR phase transitions, see Fig.~4 and Fig.~6, respectively. In the scale-free domain, $2<\gamma<3$, the ordering in the network spin system appears as a result of the interplay between magnetizations in $x$- and $z$-directions. The SR phase transition in this domain occurs for some FM state, see Fig.~4(b).

Finally, we have examined the quantum phase transition that happens in the zero temperature limit.
We have demonstrated that the transverse field amplitude $\lambda$ exhibits a familiar scaling as a function of frequency $\Omega_a$.
For vanishing $H$
the maximal value of the field is characterized by degree ${{\langle}k{\rangle}}$ and spin-spin interaction energy $J$. In fact, this phase transition disappears for the Dicke model with $J=0$.

It is remarkable that the results obtained can be useful for studying the controllability problem in the complex networks~\cite{Slotine}. The quantized (transverse) field represents an additional degree of freedom for this problem. In a more general case it is necessary to examine the limit when each node interacts with its own (local) quantized transverse field. 

\section{ACKNOWLEDGMENTS}
This work was financially supported by the Ministry of Science and Higher Education of Russian Federation, Goszadanie No. 2019-1339.
\appendix

\section{PHASE TRANSITIONS AT LOW AND HIGH TEMPERATURE LIMITS}

Let us briefly discuss the main properties of the regular networks at low temperatures $\beta\gg1$.
Eq.~\eqref{Reg_lb} in the limit of (21) may be rewritten as (cf.~\cite{Landau}) 
\[ \Omega_a \Gamma_0 =1- 2e^{-\beta\Gamma_0},~\eqno(A1)\]
where $\Gamma_0=\sqrt{({\Theta}k_0+H)^2+4{\lambda}^2}$. We assume that at low enough temperatures the spin system possesses the FM state with $S_z\simeq1$.

At critical temperature $T_c^{(1)}\equiv1/\beta_c^{(1)}$ Eq.~(A1) looks like \[ \Omega_a \Gamma_{0,c} =1- 2e^{-\beta_c^{(1)}\Gamma_{0,c}},~\eqno(A2)\]
where we define $\Gamma_{0,c}\equiv\Gamma_{0}(\lambda=0)=\Theta k_{0} +H$ at the phase transition point $\lambda=0$. Inverting~(A2) for the critical temperature of phase transition $T_c^{(1)}$ one can immediately obtain~(22).
 
On the other hand, at zero temperature Eq.~(A1) implies \[ \Omega_a \Gamma_{0,0} =1,~\eqno(A3)\]
where we denote $\Gamma_{0,0}\equiv\sqrt{\Gamma_{0,c}^2+4\lambda_0^2}$ and introduce $\lambda_0$ as a maximal value of the order parameter obtained at $T=0$.

To find $\lambda_0$ we assume that it is sufficiently small in the low temperature domain (cf.~\cite{Al1}), i.e. we suppose that $\lambda_0^2\ll1$, and for $\Gamma_{0,0}$ we can use approximation
\[ \Gamma_{0,0}\approx \Gamma_{0,c}(1+\frac{2\lambda_0^2}{\Gamma_{0,c}^2}).~\eqno(A4)\]

Substituting~(A4) into~(A3) and combining it with~(A2) after some straightforward calculations we obtain
\[\lambda_0=\Gamma_{0,c}e^{-\frac{1}{2}\beta_c^{(1)}\Gamma_{0,c}}.~\eqno(A5)\]
Then we can find the temperature dependence for order parameter $\lambda$ from Eqs.~(A1) and~(A2) leading to 
\[\frac{\Gamma_0}{\Gamma_{0,c}} \simeq \frac{1- 2e^{-\beta\Gamma_0,c}}{1- 2e^{-\beta_c^{(1)}\Gamma_0,c}}.~\eqno(A6)\]
In~ (A6) we  use ~(A4) and also assume that condition $\beta \lambda^2/\Gamma_{0,c}\ll1$ is fulfilled within the low temperature limit domain. 

Finally, after some calculations from~(A6) we obtain~(24) by means of~(A4),~(A5).
 
\bibliography{biblio} 

\providecommand{\noopsort}[1]{}\providecommand{\singleletter}[1]{#1}%
\begin{thebibliography}{10}
\expandafter\ifx\csname url\endcsname\relax
  \def\url#1{\texttt{#1}}\fi
\expandafter\ifx\csname urlprefix\endcsname\relax\def\urlprefix{URL }\fi
\expandafter\ifx\csname href\endcsname\relax
  \def\href#1#2{#2} \def\path#1{#1}\fi

\bibitem{Dorogov1}
S.~N. Dorogovtsev, Lectures on complex networks, Oxf. Master Ser. Phys 20
  (2010).

\bibitem{New1}
M.~Newman, Networks, Oxford University Press (2018) 727.

\bibitem{Barabook}
\textcolor{black}{Barabási, Albert-László}, Network science,
  \textcolor{black}{Cambridge University Press} (2016) 475.

\bibitem{Fortunato}
S.~Fortunato, D.~Hric, Community detection in networks:\textcolor{black}{ A}
  user guide, Physics Reports 659 (2016) 1--44.

\bibitem{Easley}
D.~Easley, J.~Kleinberg, Networks, crowds, and markets, Cambridge : Cambridge
  university press (2010) 727.

\bibitem{Napoli1}
\textcolor{black}{Krishnan, J. Torabi, R. Schuppert, A. and Di Napoli, E. },
  \textcolor{black}{A modified Ising model of Barabási–Albert network with
  gene-type spins}, Journal of Mathematical Biology 81 (2020) 769.

\bibitem{Ahnert}
S.~E. Ahnert, W.~P. Grant, C.~J. Pickard, Revealing and exploiting hierarchical
  material structure through complex atomic networks, npj Computational
  Materials 3~(1) (2017) 1--8.

\bibitem{Bianconi}
G.~Bianconi, Superconductor-insulator transition on annealed complex networks,
  Physical Review E 85~(6) (2012) 061113.

\bibitem{Nori}
Z.-L. Xiang, S.~Ashhab, J.~Q. You, F.~Nori, Hybrid quantum circuits:
  Superconducting circuits interacting with other quantum systems, Rev. Mod.
  Phys. 85 (2013) 623.

\bibitem{Barrat}
A.~Barrat, M.~Barthelemy, A.~Vespignani, Dynamical processes on complex
  networks, Cambridge University Press (2008) 366.

\bibitem{Brito}
S.~Brito, A.~Canabarro, R.~Chaves, D.~Cavalcanti, Statistical properties of the
  quantum internet, Physical Review Letters 124~(21) (2020) 210501.

\bibitem{Bullmore}
E.~Bullmore, O.~Sporns, Complex brain networks: graph theoretical analysis of
  structural and functional systems, Nature Reviews Neuroscience 10~(3) (2009)
  186--198.

\bibitem{Albert}
R.~Albert, A.~L. Barabási, Statistical mechanics of complex networks, Nature
  Reviews Neuroscience 74~(1) (2002) 47.

\bibitem{Dorogovtsev}
S.~Dorogovtsev, A.~Goltsev, J.~F. Mendes, Critical phenomena in complex
  networks, Reviews of Modern Physics 80~(4) (2008) 1275.

\bibitem{Holyst}
J.~A. Hołyst, K.~Kacperski, F.~Schweitzer, Phase transitions in social impact
  models of opinion formation, Physica A: Statistical Mechanics and its
  Applications 285~(1-2) (2000) 199--210.

\bibitem{Tsarev}
D.~Tsarev, A.~Trofimova, A.~Alodjants, A.~Khrennikov, Phase transitions,
  collective emotions and decision-making problem in heterogeneous social
  systems, Scientific Reports 9~(1) (2019) 1--13.

\bibitem{Thurner}
T.~Pham, I.~Kondor, R.~Hanel, S.~Thurner, The effect of social balance on
  social fragmentation, arXiv: 2005.01815v1 (2020).

\bibitem{Leone}
M.~Leone, A.~Vazquez, F.~Vespignani, R.~Zecchina, Ferromagnetic ordering in
  graphs with arbitrary degree distribution, Eur. Phys. J. B 28 (2002)
  191--197.

\bibitem{Bianconi2}
G.~Bianconi, Enhancement of $t_c$ in the superconductor–insulator phase
  transition on scale-free networks, Journal of Statistical Mechanics: Theory
  and Experiment 2012~(7) (2012) P07021.

\bibitem{Krasnytska}
M.~Krasnytska, B.~Berche, Y.~Holovatch, R.~Kenna, Partition function zeros for
  the \textcolor{black}{Ising} model on complete graphs and on annealed
  scale-free networks, Physics A: Mathematical and Theoretical 49~(13) (2016)
  135001.

\bibitem{Barabasi}
G.~Bianconi, A.~L. Barabási, \textcolor{black}{Bose-Einstein} condensation in
  complex networks, Physical Review Letters 86~(24) (2001) 5632.

\bibitem{Bianconi3}
G.~Bianconi, Mean field solution of the \textcolor{black}{Ising model on a
  Barabási–Albert} network, Physics Letters A. 303~(2-3) (2002) 166--168.

\bibitem{Suchecki}
K.~Suchecki, J.~A. Hołyst, Order, disorder and criticality: advanced problems
  of phase transition theory, World Scientific Publishing 3 (2013) 167--200.

\bibitem{Dorgov2}
S.~N. Dorogovtsev, A.~V. Goltsev, J.~F.~F. Mendes, \textcolor{black}{Ising}
  model on networks with an arbitrary distribution of connections, Physical
  Review E 66~(1) (2002) 016104.

\bibitem{Aleksiejuk}
A.~Aleksiejuk, J.~A. Hołyst, D.~Stauffer, Ferromagnetic phase transition in
  \textcolor{black}{Barabási–Albert} networks, Physica A: Statistical
  Mechanics and its Applications 310~(1-2) (2002) 260--266.

\bibitem{Napoli2}
\textcolor{black}{Krishnan, J. Torabi, R. Di Napoli, E. Honerkamp, C.
  Schuppert, A.}, \textcolor{black}{A Long-Range Ising Model of a
  Barabási-Albert Network}, arXiv:2005.05045 (2020).

\bibitem{Suzuki}
S.~Suzuki, J.~I. Inoue, B.~K. Chakrabarti, Quantum \textcolor{black}{Ising}
  phases and transitions in transverse ising models, Springer Heidelberg, New
  York, Dordrecht, London 862 (2012) 403.

\bibitem{Nori2}
I.~Buluta, S.~Ashhab, N.~Franco, Natural and artificial atoms for quantum
  computation, Rep. Prog. Phys. 74 (2011) 104401.

\bibitem{Hepp}
K.~Hepp, E.~H. Lieb, Equilibrium statistical mechanics of matter interacting
  with the quantized radiation field, Physical Review A. 8~(5) (1973) 2517.

\bibitem{Wang}
Y.~K. Wang, F.~T. Hioe, Phase transition in the \textcolor{black}{Dicke} model
  of superradiance, Physical Review A 7~(3) (1973) 831.

\bibitem{Emary}
C.~Emary, T.~Brandes, Chaos and the quantum phase transition in the
  \textcolor{black}{Dicke} model, Physical Review E 67~(6) (2003) 066203.

\bibitem{Larson}
J.~Larson, E.~K. Irish, Some remarks on ‘superradiant’phase transitions in
  light-matter systems, Journal of Physics A: Mathematical and Theoretical
  50~(17) (2017) 174002.

\bibitem{Bohnet}
J.~G. Bohnet, Z.~Chen, M.~D. Weiner, J.~M., M.~J. Holland, J.~K. Thompson, A
  steady-state superradiant laser with less than one intracavity photon, Nature
  484~(7392) (2017) 7392.

\bibitem{Chestnov}
I.~Y. Chestnov, A.~P. Alodjants, S.~M. Arakelian, Lasing and high-temperature
  phase transitions in atomic systems with dressed-state polaritons, Physical
  Review A 88~(6) (2013) 063834.

\bibitem{Akkermans}
E.~Akkermans, A.~Gero, R.~Kaiser, Photon localization and
  \textcolor{black}{Dicke} superradiance in atomic gases, Physical Review
  Letters 101~(10) (2008) 103602.

\bibitem{Eastham}
P.~R. Eastham, P.~B. Littlewood, Bose condensation of cavity polaritons beyond
  the linear regime: The thermal equilibrium of a model microcavity, Physical
  Review B. 64~(23) (2001) 235101.

\bibitem{Wang2}
Z.~Wang, H.~F.~W. Li, X.~Song, C.~Song, W.~Liu, H.~Wang, Controllable switching
  between superradiant and subradiant states in a 10-qubit superconducting
  circuit, Physical Review Letters 124~(1) (2020) 013601.

\bibitem{Bamba}
M.~Bamba, K.~Inomata, Y.~Nakamura, Superradiant phase transition in a
  superconducting circuit in thermal equilibrium, Physical Review Letters
  117~(17) (2016) 173601.

\bibitem{Cong}
K.~Cong, Q.~Zhang, Y.~Wang, G.~T. Noe, A.~Belyanin, J.~Kono, Dicke
  superradiance in solids, JOSA B 33~(7) (2016) 80--101.

\bibitem{Yalouz}
S.~Yalouz, P.~V., Continuous-time quantum walk on an extended star graph:
  Trapping and superradiance transition, Physical Review E 97~(2) (2018)
  022304.

\bibitem{Vahala}
K.~Vahala, Optical microcavities, Nature 424~(6950) (2003) 839--846.

\bibitem{Al1}
A.~Alodjants, I.~Barinov, S.~Arakelian, Strongly localized polaritons in an
  array of trapped two-level atoms interacting with a light field, Journal of
  Physics B: Atomic, Molecular and Optical Physics 43~(9) (2010) 095502.

\bibitem{Yamamoto}
H.~Deng, H.~Haug, Y.~Yamamoto, Exciton-polariton
  \textcolor{black}{Bose-Einstein} condensation, Rev. Mod. Phys. 82 (2010)
  1589.

\bibitem{Chest1}
I.~Chestnov, A.~Alodjants, S.~Arakelian, J.~Klaers, F.~Vewinger, M.~Weitz,
  \textcolor{black}{Bose-Einstein} condensation for trapped atomic polaritons
  in a biconical waveguide cavity, Physical Review A 85~(5) (2012) 053648.

\bibitem{Ber}
O.~Berman, Y.~Lozovik, D.~Snoke, Theory of \textcolor{black}{Bose-Einstein}
  condensation and superfluidity of two-dimensional polaritons in an in-plane
  harmonic potential, Physical Review B 77~(15) (2008) 155317.

\bibitem{Gammelmark}
S.~Gammelmark, K.~Mølmer, Phase transitions and heisenberg limited metrology
  in an\textcolor{black}{ Ising} chain interacting with a single-mode cavity
  field, New Journal of Physics 13~(5) (2011) 053035.

\bibitem{Lee}
C.~F. Lee, N.~F. Johnson, First-order superradiant phase transitions in a
  multiqubit cavity system, Physical Review Letters 93~(8) (2004) 083001.

\bibitem{Bazhenov}
A.~Y. Bazhenov, D.~V. Tsarev, A.~P. Alodjants, Temperature quantum sensor on
  superradiant phase-transition, Physica B: Condensed Matter. 579 (2020)
  411879.

\bibitem{Khrennik1}
A.~Khrennikov, Social laser’: action amplification by stimulated emission of
  social energy, Philosophical Transactions of the Royal Society A:
  Mathematical, Physical and Engineering Sciences 374~(2058) (2016) 20150094.

\bibitem{Khrennik2}
A.~Khrennikov, Z.~Toffano, F.~Dubois, Concept of information laser: from
  quantum theory to behavioural dynamics, The European Physical Journal Special
  Topics 227~(15) (2019) 2133--2153.

\bibitem{Cota2}
W.~Cota, S.~Ferreira, R.~Pastor-Satorras, Quantifying echo chamber effects in
  information spreading over political communication networks, EPJ Data Science
  8~(1) (2019) 35.

\bibitem{Baumann1}
F.~Baumann, P.~Lorenz-Spreen, I.~Sokolov, M.~Starnini, Modeling echo chambers
  and polarization dynamics in social networks, Physical Review Letters 124~(4)
  (2020) 048301.

\bibitem{Bukh}
V.~Guleva, E.~Shikov, K.~Bochenina, K.~S., A.~Alodjants, A.~Boukhanovsky,
  Emerging complexity in distributed intelligent systems, Entropy 22 (2020)
  1437.

\bibitem{Lee2}
S.~H. Lee, M.~Ha, H.~Jeong, J.~D. Noh, H.~Park, Critical behavior of the
  \textcolor{black}{Ising} model in annealed scale-free networks, Physical
  Review E 80~(5) (2009) 051127.

\bibitem{Landau}
E.~Lifshitz, L.~Pitaevskii, Statistical physics: Theory of the condensed state,
  part 2, Elsevier (2013) 38.

\bibitem{SnokeL}
D.~Snoke, \textcolor{black}{ Polariton Condensation and Lasing},
  \textcolor{black}{In: Timofeev V., Sanvitto D. (eds) Exciton Polaritons in
  Microcavities. Springer Series in Solid-State Sciences, Springer, Berlin,
  Heidelberg} 172 (2012) 307.

\bibitem{Slotine}
Y.-Y. Liu, J.~Slotine, A.~Barabasi, Controllability of complex networks, Nature
  473 (2011) 167.

\end{thebibliography}
\end{document}